\documentclass[preprint2]{aastex}

\newcommand{\etal}{et~al.~}
\newcommand{\kms}{km~s$^{-1}$}
\slugcomment{To appear in ApJ, 2008} %\today} %Not to appear in Nonlearned J., 45.}

\shorttitle{The Globular Cluster System of M60}
\shortauthors{Lee \etal}

\begin{document}


\title{Washington CCD Photometry of the Globular Cluster System of the Giant Elliptical Galaxy M60 in Virgo}


\author{Myung Gyoon Lee\altaffilmark{1},  Hong Soo Park\altaffilmark{1}, 
 Eunhyeuk Kim\altaffilmark{1}, Ho Seong Hwang\altaffilmark{1,2}, Sang Chul Kim\altaffilmark{3}, 
Doug Geisler\altaffilmark{4}}

\author{To appear in ApJ, 2008}
%\affil{Astronomy Department, University of California,
%    Berkeley, CA 94720}

%\author{C. D. Biemesderfer\altaffilmark{4,5}}
%\affil{National Optical Astronomy Observatories, Tucson, AZ 85719}
%\email{aastex-help@aas.org}

%\and

%\author{R. J. Hanisch\altaffilmark{5}}
%\affil{Space Telescope Science Institute, Baltimore, MD 21218}

\altaffiltext{1}{Astronomy Program, Department of Physics and Astronomy,
Seoul National University, Seoul 151-742,
Korea; mglee@astro.snu.ac.kr}
\altaffiltext{2}{Current address: Korea Institute for Advanced Study, Seoul, Korea}
\altaffiltext{3}{Korea Astronomy and Space Science Institute,
Daejeon 305-348, Korea}
%\altaffiltext{3}{Center for Astrophysics, ..}
\altaffiltext{4}{Departamento de Fisica,  Universidad de Concepci\'on, 
Casilla 160-C, Concepci\'on, Chile}

%\altaffiltext{1}{Visiting Astronomer, Cerro Tololo Inter-American Observatory.
%CTIO is operated by AURA, Inc.\ under contract to the National Science
%\altaffiltext{2}{Society of Fellows, Harvard University.}
%\altaffiltext{3}{present address: Center for Astrophysics,
 %   60 Garden Street, Cambridge, MA 02138}
%\altaffiltext{4}{Visiting Programmer, Space Telescope Science Institute}
%\altaffiltext{5}{Patron, Alonso's Bar and Grill}


\begin{abstract}
We present a photometric study of the globular clusters 
in the giant elliptical galaxy M60 %(NGC 4649) 
in the Virgo cluster, based on deep, relatively  wide field %($16\arcmin.4 \times16\arcmin.4$ ) 
Washington $CT_1$ CCD images. 
The color-magnitude diagram reveals a significant population 
of globular clusters in M60, and a large number of young luminous clusters
in NGC 4647, a small companion spiral galaxy north-west of M60.
The color distribution of the globular clusters in M60 is clearly bimodal,
with a blue peak at $(C-T_1)=1.37$, and a red peak at $(C-T_1)=1.87$.
We derive two new transformation relations between the $(C-T_1 )_0$ color and [Fe/H]
using the data for the globular clusters in our Galaxy and M49.
Using these relations we derive the metallicity distribution of the globular clusters in M60,
which is also bimodal: 
a dominant metal-poor component with center at [Fe/H]$ = -1.2$, 
and a weaker metal-rich component with center at [Fe/H]$ = -0.2$.
The radial number density profile of the globular clusters is 
more extended than that of the stellar halo, and the radial number density profile of
the blue globular clusters is more extended than that of the red globular clusters.
The number density maps of the globular clusters show that
the spatial distribution of the blue globular clusters is roughly circular,
while that of the red globular cluster is elongated similarly to that of the 
stellar halo.
We estimate the total number of the globular clusters in M60 to be $3600\pm500$,
and the specific frequency to be $S_N=3.8\pm0.4$.
The mean color of the bright blue globular clusters gets redder as they get brighter
in both the inner and outer region of M60. 
This blue tilt is seen also in the outer region of M49, the brightest Virgo galaxy.
Implications of these results are discussed.
\end{abstract}




\section{Introduction}

Old globular clusters keep the fossil record for the early epoch of their
host galaxies as well as globular clusters themselves.
By studying the age and metallicity of these globular clusters we can investigate 
the formation and early evolution of their host galaxies as well as globular clusters themselves.
Globular clusters are distributed in a much wider region than the halo stars in
their host galaxy,
and thousands of them are found in giant elliptical galaxies (gEs).
%These features, combined with their high luminosities, unique and 
This feature, combined with easily-derived velocity, 
makes globular
clusters powerful probes with which to study the structure and
kinematics in the outer halo of nearby gEs as well as the inner
regions (see \citet{lee03} and \citet{bro06}, and references therein).

While the globular clusters in the inner region of nearby gEs have been
 extensively studied using the Hubble Space Telescope (HST) 
\citep{kun01, lar01, pen06, mie06, str06, har06, jor07}, 
the globular clusters in the outer halo of nearby gEs have been studied using the wide field camera in the ground-based telescopes
\citep{lee93,mcl93, gei96b,lee98,rho01,rho04,dir04, har04, tam06a, tam06b, bas06}.
However, the number of gEs for which the globular clusters were studied 
using the wide field camera is still small.

Virgo, the nearest galaxy cluster, is one of the best targets for the study of
globular cluster in gEs, because 
it includes several gEs that are almost at the same distance from us.
We have been carrying a long-term photometric study of globular clusters in Virgo gEs using the
Washington filter system \citep{lee93,gei96,lee98}.
The Washinton system is known to be very sensitive to measuring the metallicity of the globular clusters and has a wide bandwidth so that it is ideal for studying the metallicity of the extragalactic globular clusters \citep{gei90}.
The results for the globular clusters in M87, the central cD in Virgo, were given in \citet{lee93},
and those for the globular clusters in M49, the brightest gEs in Virgo, were given in \citet{gei96b,lee98}.
Here we present the results for the third Virgo gE, M60 in this series.
M60 was selected because
it is one of the brightest gEs in Virgo and is therefore expected to
possess a rich globular cluster system.

M60 (NGC 4649) is only slightly less luminous ($M_V=-22.44$ mag) %-22.14$ mag) 
than two brightest Virgo galaxies 
M87 ($M_V=-22.62$ mag) %-22.38$ mag) 
and M49 ($M_V=-22.83$ mag). %-22.57$ mag).
M60 is of morphological type E2, and is ultraviolet-bright, emitting strong flux at $\lambda <2500 $\AA  ~\citep{ber82}. A recent Chadra image of M60 shows that
the diffuse X-ray emission is detected out to about $3\arcmin$ from the center of M60,
with a circular shape \citep{hum06}.
Basic information on M60 is listed in Table 1.
We adopted a distance to M60, % 16.83 Mpc ($(m-M)_0=31.13\pm 0.15$)  ton01}, of
of 17.30 Mpc ($(m-M)_0=31.19\pm 0.07$)   %$(m-M)_V=31.19+0.088=31.278 E(V-I)=0.037$
based on the surface brightness fluctuation method in \citet{mei07} 
for which one arcsec corresponds to 84 pc. %81 pc.
Foreground reddening toward M60 is very small, $E(B-V)=0.026$ \citep{sch98},
corresponding to $E(C-T_1)=1.966 E(B-V)=0.051$, $A(T_1)=0.071$, and $A(V)=0.088$.

M60 has a companion SBc galaxy, NGC 4647, 
located  $2.\arcmin5$ from the center of M60 in the north-west
direction (corresponding to a projected distance of $\sim$12.6 kpc for the adopted distance to M60).
The radial velocity of NGC 4647 ($v=1422$ \kms) 
is 305~ \kms ~larger than that of M60 ($v=1117$ \kms).
\citet{whi00} found no foreground absorption due to NGC 4647 in the area of M60, 
and were unable to tell whether  M60 or NGC 4647 is closer.  

\citet{cou91} performed the first photometric study based on $BV$ CCD imaging 
of the globular clusters in a small  field ($2\arcmin.1 \times 3\arcmin.4$) of M60. 
They found a large dispersion in color distribution, and a radial gradient in
the mean cluster colors. They also found that the mean  color of the globular clusters in M60 ($(B-V)=0.75$) is 0.1 mag redder than that of M49 ($(B-V)=0.65$). 
\citet{har91} determined the $B$ band luminosity function 
up to $B\sim$26 mag of the globular clusters in this field,
and found that the globular clusters follow a more extended spatial distribution than the stellar light. 
Later, photometric studies based on HST/WFPC2  as well as ground-based images
revealed that the M60 globular cluster system has a %distinct 
clear bimodality in
the color distribution \citep{nei99,kun01,lar01,for04}. 
\citet{for04} found from the analysis of wide field images
obtained using Gemini/GMOS 
that the red globular clusters have a similar surface density
distribution to that of the stellar light, which is steeper than that
of the blue globular clusters. In addition, they derived a value for the  specific
frequency of the M60 globular clusters of $S_N\sim4.1\pm1.0$, which is lower than
the value $S_N\sim6.7$ given in \citet{ash98} based on inferior data.

Recently, it has been found that M60 is one of the galaxies that show a ``blue-tilt''
in the color-magnitude diagram of its  globular clusters, in which
the brighter the blue globular clusters are, the redder they are in the mean \citep{har06, str06, mie06}.
\citet{sar03} and \citet{ran04} detected some discrete sources in M60 in the X-ray band
using Chandra, and \citet{ran04} found that roughly 47\% of the X-ray discrete
sources is identified with globular clusters.
By cross-correlating Chandra point sources and optical globular cluster candidates,
\citet{kim06} found that the mean probability for a globular cluster to harbor a
low mass X-ray binary (LMXB) in M60 is about 6.1$\pm$1.0 \%, and that
this probability for the red GCs is much larger than that for the blue globular clusters. 
%An analysis of the structure and mass profile of M60  derived from the Chandra images was given by \citet{hum06}.
In addition, \citet{pie06} published a spectroscopic
study of  38 globular clusters in M60 based on data obtained using Gemini/GMOS,
and \citet{bri06} presented a study of the globular cluster kinematics of M60
using these data.

In this paper we present a photometric study of globular clusters in M60
using deep, relatively wide field CCD photometry.
This study covers a field of M60 that is much larger than any previous studies 
on globular clusters in M60, and is using the Washington system that has a better sensitivity
for measuring the metallicity compared with most other systems.
This study supplements also our kinematic studies of the globular clusters in M60 \citep{lee07, hwa07}.
This paper is organized as follows. 
In Section 2 we describe our observations and data reduction. 
In \S 3 we present the color-magnitude diagram and color distribution of
the globular clusters, 
and compare the surface photometry of the stellar halo with the structure
of the globular cluster system. 
We also investigate the radial variation of the mean magnitudes
and colors of the globular clusters, and estimate the total number and specific frequency
of the globular clusters.
In \S 4 we discuss our results and their implication in comparison with other
studies. 
Primary results are summarized in the final section.

\section{Observations and Data Reduction}

\subsection{Observations}

CCD images of M60 were obtained on the photometric nights of April 9 and 10, 1997 (UT)
at the KPNO 4 m telescope using the $2048 \times 2048$ pixels CCD camera at 
the prime focus.
Washington $C$ and Kron-Cousins $R$ filters were used. The 
Kron-Cousins $R$ filter has a very similar effective wavelength to that of the Washington $T_1$ filter, 
but with a much wider bandwidth resulting in three times the    sensitivity \citep{gei96}. \citet{gei96} showed that $T_1$ reproduces 
$R$ magnitude very well:
$R=T_1 + 0.003-0.017 (C-T_1 )$ with rms=0.02.
So we used the $R$ filter as an alternative to the $T_1$ filter, as done in
most recent Washington studies \citep{gei96, lee98, dir04, har04, bas06}.

%We call this $R$ filter as $T_1$ filter in the rest of this paper.

The observation log is given in Table 2.
The size of the field of view is $16\arcmin.4 \times 16\arcmin.4$, and the pixel scale
is 0.47 arcsec pixel$^{-1}$. Exposure times are 100 s and %??this is 4 not 5???
$4 \times 1500$ s for $C$,
and 60 s and $3 \times 1000$ s for $T_1$.
Each long exposure image was taken with dithering of 10 to 28 arcsec.
The seeing ranged from 1.1 to 1.5 arcsec.
In addition, several Washington standard fields in \citet{gei96} 
were observed during the observing run.

\subsection{Point Source Photometry}

Each frame was trimmed, bias-subtracted and flat-fielded with twilight skyflats
for $C$ and dome flats for $T_1$ using the IRAF software.
%Accuracy of the flat field is estimated to be smaller than 2????????\%.
The individual long exposure images in each filter were then shifted to a common center and
medianed together.
Figure 1 displays a grayscale map of the short exposure $T_1$ image of M60.

Globular clusters at the distance of M60 appear as point sources in our images. 
First we subtracted the stellar halo of M60 from each of the original images for better detection of the sources as follows.
We created a model image of M60 using the
ELLIPSE task in IRAF/STSDAS. Then we subtracted the model
image from the original image to remove the stellar halo light, and then added
a constant equal to the mean value of the sky background in the original image.
We used the resulting images for the detection of objects with 
the digital photometry program DAOPHOT/ALLFRAME \citep{ste94}.
We used threshold=3$\sigma$, sharpness=0.2--1.0, roundness=-0.7 to 0.7
 for detection in the KPNO images.
Instrumental magnitudes of the detected objects 
in the images were derived using the point-spread
function (PSF)-fitting routine in DAOPHOT/ALLFRAME. 
To take care of the variation of the PSF we adopted a quadratically variable PSF, and applied the aperture correction to the PSF-fitting magnitudes.
The value of the aperture correction was derived from the difference between
the aperture magnitudes and PSF-fitting magnitudes for 
isolated bright stars:
0.014  and 0.008 for short and long exposure $C$ images, respectively,
and 0.033 and --0.009 for short and long exposure $T_1$ images, respectively.

The list of detected objects includes both point sources and extended sources.
We selected the point sources among the detected sources using the morphological classifier $r_{-2}$  moment defined 
as $r_{-2} = [{ \Sigma_i (I_i -sky) / \Sigma_i [(I_i -sky ) / (r_i^2 + 0.5)] ]}^{1/2}$, 
where $r_i$ is the radial distance of the i-th pixel from the center
of the source, 
$I_i$ is the intensity value at the i-th pixel, and {\it sky} is the local
background value derived from a more distant annulus \citep{kro80}.
The FWHMs of the point sources vary approximately as a quadratic function of the radius from the
center of the image. So  $r_{-2}$ was normalized to the value for the center of the image.
From the artificial star experiment to be described below, we decided to consider
the sources with $r_{-2} <1.27$ to be point sources.
Misclassification rates are negligible for the bright objects with $T_1<23$ and
increase with magnitude, reaching about 20\% at $T_1=24$. 
The central regions at $r\lesssim1$ arcmin were saturated in the long exposure images so
that we used the photometry of the objects in the central region derived from the short exposure images.

For this reason, 
the KPNO images are of limited use for the study of the globular clusters in the central region of M60. 
Therefore we analyzed F555W and F814W images of the central region  ($2\arcmin \times 2\arcmin$) of M60  in the HST/WFPC2 archive to supplement the KPNO data. Hereafter we call $V$ and $I$ for F555W and F814W, respectively.
The position of the HST/WFPC2 field is marked in Figure 1 
(see \citet{kim06} for the journal of its observation). 
We created a model image of M60 using the
ELLIPSE task in IRAF/STSDAS. Then we subtracted the model
image from the original image to remove the stellar halo light.
We detected objects in the WFPC2 images and derived the photometry of the detected objects
 using the digital photometry software HSTPHOT \citep{dol00}. 
We used 3.5 $\sigma$ as a threshold for detection in the HST images.
We used the radii of the aperture
of 3 pixels for PC chips and 2 pixels for WF chips to get the aperture magnitudes of the detected objects, and
applied the aperture correction given by \citet{kun01} to get the total magnitudes
of the detected objects.
We selected the starlike sources in the list of the objects returned by HSTPHOT using $r_{-2}$.
We used the HST photometry for the analysis of the central region ($r<1\arcmin.5$) on one side of M60 
for which the KPNO photometry is poor.
We applied the same procedure  to the images of another HST/WFPC2 field, at $5\arcmin$ north
from the center of M60 (called the north HST/WFPC2 field), as marked in Figure 1
(see \citet{kim06} for the basic information of this field).
It is found that the number of globular clusters in this field is  so small that they are of limited
use for this study. 
 
\subsection{Standard Calibration}

We derived the transformation equations for standard calibration from the photometry of the 
Washington standard stars observed during the same night.
We obtained the instrumental magnitudes of the standard stars 
using the aperture 
radius   of 7.5 arcsec as used in \citet{gei96}.
The standard transformation equations we derived are:
(April 9) 
$T_1 = t_1 + 0.034 (c-t_1) -0.122 X+ 0.282 $ 
with rms=0.021 and N=42, and
$(C-T_1) = 1.062 (c-t_1) -0.263 X - 0.405 $ 
with rms=0.018 and N=44, 
and (April 10 ) 
$T_1 = t_1 + 0.034 (c-t_1) -0.122 X + 0.269 $ with rms=0.028 and N=39, and
$(C-T_1) = 1.062 (c-t_1) -0.263 X - 0.425 $ with rms=0.030 and N=39,
where the upper case letters represent the standard magnitudes, the lowercase letters
the instrumental magnitudes (with a DAOPHOT system
zero point of 25.0), and $X$ the air mass.
We transformed the instrumental magnitudes of the sources onto the standard
system using these transformation equations.
We selected and listed in Table 3 the $CT_1$ photometry of 4497 point sources with $\sigma(C-T_1 ) <0.3$
measured in the KPNO images, and
Figure 2 displays the mean errors of $T_1$ and $(C-T_1)$ versus $T_1$ magnitude.

\subsection{Completeness of the Photometry}

We estimated the completeness of the KPNO photometry using DAOPHOT/ADDSTAR  
that was designed for the artificial star experiment.
We generated a set of artificial stars for which the color-magnitude diagrams and luminosity functions are similar to the observational one, using the PSFs derived from the long exposure real images. 
We did not include the central region at $r<1\arcmin$ that was saturated in the real images.
Then we added them to the real image avoiding the position of detected real objects. 
We added 1200 artificial stars to each pair of $C$ and $T_1$ images in a set of
50 pairs so that the total number of added artificial stars is 60,000.
Then we applied the same procedure of photometry to the artificial images as used for the real images,
and estimated as the completeness factor, the number ratio of the recovered artificial stars
and the added artificial stars.

Figure 3 displays the completeness for $C$ and $T_1$ we derived for the point sources.
We plotted the completeness for the entire region and five bins in radial distance:
$1\arcmin<r<2\arcmin$, $2\arcmin<r<3\arcmin$,  $3\arcmin<r<4\arcmin$, $4\arcmin<r<5\arcmin$, and $5\arcmin<r<6\arcmin$.
It is seen that 
the completeness is higher than 90 \% for $T_1=23$  ($C=24.3$) for the outer region at $r>2\arcmin$ and $T_1=22.8$ ($C=24.0$) for the inner region at $1\arcmin<r<2\arcmin$.
The photometric limit levels for 50 \% completeness are $T_1=24.4$  ($C=25.4$) for the outer region at $r>2\arcmin$ and $T_1=23.9$ ($C=25.0$) for the inner region 
at $1\arcmin<r<2\arcmin$.
The completeness varies little depending on radius for the outer region at $r>2\arcmin$, 
while it is somewhat lower for the inner region at $r<2\arcmin$ compared with the outer region.
We derived the mean photometric errors from the difference between the magnitudes
of the added objects and recovered objects, finding that they are similar to the
measured photometric errors for the real objects as seen in Figure 2.
We derived similarly the completeness for  $V$ and $I$ using HSTPHOT, which is plotted also
in Figure 3. The completeness for the HST photometry is higher than to 95 \% for $V<23.5$ mag for all the radial range of the HST data.
 
\subsection{Comparison with Previous Photometry}

There is no previous Washington photometry of the sources in M60. However there
are a few photometry of the sources in M60 using the different filter systems in the
literature. We derive the transformation relations between our Washington photometry
and the $gi$ photometry given by \citet{for04}.
%so  we compared our photometry with that  obtained with other filter systems available in the literature.

\citet{for04} presented $gi$ photometry of the globular clusters in a   field of about 90 square arcmin including M60, based on the CCD images taken using the Gemini North telescope.
Our field of view covering 256 square arcmin is about 2.8 times larger than that of \citet{for04}. 
We compared ($C-T_1$) colors and ($g-i$) colors for 396 objects with $T_1<22$ mag in common between this study and \citet{for04}, as displayed in Figure 4.
Figure 4 shows that the relation between ($g-i$) colors and ($C-T_1$) colors can be fit 
by a triple linear relation as follows:
$(g-i)=0.735 (C-T_1 ) - 0.121$ with rms=0.110 for $0.2<(C-T_1 )<1.2$,
$(g-i)=0.527 (C-T_1 ) + 0.119$ with rms=0.092  for $1.2<(C-T_1 )<2.4$, and
$(g-i)=1.372 (C-T_1 ) - 1.943$ with rms=0.243  for $2.4<(C-T_1 )<4$.
In addition the transformations between magnitudes are
derived to be: $g= C +0.340-0.548 (C-T_1 )$ with rms=0.067,
and $i= T_1 +0.429-0.199 (C-T_1 )$ with rms=0.070.

We also derived the transformation between ($C-T_1$) colors and ($V-I$) colors 
for the objects with $T_1 <22$ mag in common between the KPNO images 
and HST/WFPC2 images, as shown in Figure 4.
They show a linear relation for the color range of $0.9<(C-T_1 )<2.4$.
Linear fitting to the data for  72 objects with small errors ($\sigma(C-T_1 ) < 0.1$ ) yields
 $(V-I)=0.407 (C-T_1 ) +0.459$ with rms=0.047  for  $0.9<(C-T_1 )<2.4$.
This is in very good agreement with the result derived from the photometry
of M49 (NGC 4472) by \citet{lee00}, 
$(V-I)=0.443 (C-T_1 ) +0.396$ for $(C-T_1 )<2.6$.
We also derived an approximate relation for the magnitudes,
$V=T_1 -0.106 + 0.243 (C- T_1 )$ with rms=0.100 for 66 objects
 with $0.9<(C- T_1 )<2.4$.

\section{Results}

\subsection{Color-Magnitude Diagram}

Figure 5 displays the color-magnitude diagrams of the measured point sources 
in the KPNO images of M60 
(in three regions at $1\arcmin.5<r<7\arcmin$, $7\arcmin<r<9\arcmin$, and $9\arcmin < r < 10\arcmin$ where
$r$ is the projected galactocentric distance) as well as those in the HST/WFPC2 images (for $r<1\arcmin.5$).
We excluded the objects within a circular region of radial distance of $1\arcmin$ from the center of NGC 4647 from the KPNO data in Figure 5.
We transformed the ($V-I$) colors of the objects in the HST WFPC2 images into ($C-T_1$) colors using the transformation equation in the previous section.

Three general kinds of objects are seen in Figure 5:
(a) globular clusters of M60 residing in the broad vertical feature
in the color range of $1.0<(C-T_1)<2.4$,
(b) a small number of bright foreground stars 
at $(C-T_1) \lesssim 0.9$ and $(C-T_1) \gtrsim 2.6$,
and  
(c) faint blue unresolved background galaxies with $T_1>23$ mag. 
%There are seen two distinguishable components in the 
Two vertical features are visible in Figure 5(b) %of globular clusters 
at $1\arcmin.5<r<7\arcmin$: 
blue globular clusters (BGCs) and red globular clusters (RGCs).
We selected a sample of bright globular clusters with $1.0<(C-T_1)<2.4$  and 
$19<T_1 <23$ ($-12.26 <M_{T_1}<-8.26$) for the analysis, 
considering the following points:
1) the color range of the known globular clusters in other galaxies is 
$(C-T_1 )_0 \approx 1$ to 2.4, and the foreground reddening for M60 is small;
2) the peak luminosity of the globular clusters
is estimated to be somewhat fainter than $T\approx 23$ at the distance of M60
(see also Section 3.7);
3) the mean photometric error is smaller than $\sigma(T_1 ) = 0.05$ and the incompleteness in our photometry is estimated to be minor for $T_1 <23$; and
4) the contamination due to background galaxies is estimated to be very small
for $T_1 <23$.
The sample of globular clusters is separated around $(C-T_1)=1.7$, 
since the color boundary between the two components is estimated to be at  
$(C-T_1)=1.7$ as described in the following section.
So we divided the entire sample of globular clusters into two classes:
the BGCs with $1.0<(C-T_1)<1.7$ 
and the RGCs with $1.7<(C-T_1)<2.4$.
A small number of the BGCs are still seen even at $9\arcmin<r<10\arcmin$, while very
few RGCs are visible in the same region. %at $r>9\arcmin$.

Figure 6 displays the color-magnitude diagrams of 
the point sources % and slightly extended sources 
within a radius of one arcmin from the center of the companion spiral galaxy NGC 
4647 (upper panel) % ??? no problems with central saturation???
and in a ``control field'' with the same area (lower panel).
The region for NGC 4647 is about $2.5\arcmin$ from the center of M60 so that it
is unaffected by the saturation of the M60 nucleus in the image.
The control field is located at the same radial distance from the center of M60, 
but in the opposite direction. 
We also checked the color-magnitude diagrams for
two other control fields at the same radial distance but at different position angle
(45 degrees and 315 degrees, respectively).
They are very similar to the color-magnitude diagram for the chosen control field in Figure 6, so that the chosen control field can be considered to represent a control field for the
statistical analysis. 
The number of objects within the boundary of bright globular clusters in the field of NGC 4647 (marked by the two rectangles) 
is slightly smaller than that in the control field,
showing that most of the objects within the rectangles 
are probably globular clusters belonging to M60.
Most of the point sources in NGC 4647 are much bluer than globular clusters, 
but as bright as globular clusters of M60.

There are seen many more sources that are slightly extended, 
compared with the point sources, in NGC 4647. 
We selected the extended sources using the criterion $1.27 \le r_{-2} < 2$
(note $r_{-2} < 1.27$ for the point sources).
They are slightly larger than the point sources and were included in the output list
of the DAOPHOT/ALLFRAME. They have larger values for $\chi$ and sharpness than the
point sources, indicating that their PSF-fitting magnitudes are less reliable than those of the point sources. Although  their $r_{-2} $ values show that they are slightly larger than the point sources, it is not easy to estimate their size in our images.
We derived the photometry of these objects, plotting them in Figure 6.
%(filled squares for $r<0\arcmin.5$ and open squares for $0\arcmin.5 < r<1\arcmin$).
Most of these slightly extended sources are also much bluer 
than globular clusters, but as bright as globular clusters of M60.
These blue objects are not seen in the control field. 
Therefore these point sources and slightly extended sources with blue colors 
are considered to be young luminous star clusters or 
associations in NGC 4647. 

Recent studies show several evidences of interaction between NGC 4647 and M60,
although \citet{san94} described that the two galaxies are not interacting: 
(1) The morphology of NGC 4647 is clearly asymmetric \citep{koo01};
(2) The inner region of M60 has a strong rotational support in contrast to other
giant elliptical galaxies, and has an asymmetric rotation curve
\citep{pin03,deb01}; and 
(3) there is seen a filament extending to the
northeastern edge of M60 in the X-ray image \citep{ran06}. 
Therefore we suggest that the young clusters in NGC4647 might have formed during a recent interaction of NGC 4647 with M60.

\subsection{Color Distribution of the Globular Clusters}

Figure 7 displays the color distribution of the bright point sources 
with $19<T_1 <23$ at $1\arcmin <r< 7 \arcmin$ 
(most of which are globular clusters) in the KPNO images. 
We did not include the objects at $r<1\arcmin$ 
because they suffer from incompleteness problem.
We also derived the color distribution of the background objects with the same
magnitude range at $9\arcmin<r<10\arcmin$, finding the contribution of
these background objects is negligible as shown in Figure 7.
%subtracted it from the original color distribution for $1\arcmin <r< 7 \arcmin$, 
%and plotted the resulting net color distribution in Figure 7.
 
The color distribution is clearly bimodal.
KMM tests of the data show that the probability
that the color distribution is  bimodal over unimodal is  higher than 99.9 \%. 
We determined the following
parameters for the best-fit double Gaussian curves describing the color
distribution, using
a maximum-likelihood method through the KMM mixture modeling routine \citep{ash94}:
a primary component with center at $(C-T_1 )=1.37$ and width $\sigma=0.16$ and %width(FWHM) of 0.38, and
a secondary component with center at $(C-T_1 )=1.87$ and $\sigma=0.23$. %width of 0.55. 
We did not assume equal dispersions for the blue and red globular subpopulations
in these KMM fits and they do indeed appear to be significantly different.

The minimum between the two components is found to be at $(C-T_1)=1.7$, which was used
for dividing the entire sample into BGCs and RGCs.
This boundary color is slightly redder than that used for M49 \citep{gei96b,lee98}, but is 
consistent with \citet{cou91}'s finding based on $BV$ photometry of globular clusters in M60.
They also found that the mean  color of the globular clusters in M60 ($(B-V)=0.75$) is 0.1 redder than that of M49 ($(B-V)=0.65$). 

We investigate the radial variation of the color distribution of the bright globular clusters
after subtracting the background level derived from the region at $9\arcmin<r<10\arcmin$
in Figure 8. 
%Figure 8 shows that the color distribution of the globular clusters              varies
%with   galactocentric radius.
In the central region at $r<75\arcsec$, the RGC component looks stronger than the 
BGC component, while the BGC component begins to dominate for $r>2\arcmin$
and steadily strengthens its domination in the outer region.
The RGC component is barely seen in the bacgkround region at $9\arcmin<r<10\arcmin$.

We also derived the variation of the ratio of the  number of the bright BGCs to 
the number of bright RGCs with $r<7\arcmin$, plotting it in Figure 9.
Figure 9 shows that this ratio keeps increasing as the galactocentric radii increase,
and is fit well by the linear relation $N(BGC)/N(RGC) = 0.43(\pm0.04) r + 0.47$.
This ratio becomes unity at $r=74\arcsec$.

\subsection{Galaxy Surface Photometry}

We derived the surface photometry of M60 from the KPNO images 
using the ELLIPSE task in IRAF/STSDAS.
First we masked out bright foreground stars and nearby galaxies including NGC 4647
in the original images. 
Then we applied ELLIPSE for ellipse fitting of the isophotes of M60
to the resulting images.
We used the very outer region at $r\sim 10\arcmin$ in the corner of the image to estimate the background level.
We measured the median intensity value for each of three regions in the corner 
and took the median of the three meadian values as the background level. 
The central part ($r<3$ arcsec) of the short exposure $T_1$ image of M60 
was saturated so that we could not derive the color for this region.
The radial profiles of the surface color were obtained using the same structural 
parameters for both $C$ and $T_1$, which were derived from the $T_1$ images.
The errors for the surface brightness magnitudes are those that are given by ELLIPSE, 
not including the error for the background estimate.

Figure 10 displays the radial profiles of the surface brightness magnitudes, $(C-T_1)$ color,
 ellipticity
and position angle (PA) of M60 as a function of major radius $r_{maj}$.
We fit the surface brightness profiles with a de Vaucouleurs  $r^{1/4}$ law for the range of 
$3\arcsec.8 <r_{maj}<410\arcsec.1$, using  linear least squares fitting:
$\mu(C)= 2.568(\pm 0.011) r^{1/4} + 14.955(\pm 0.032)$ with rms=0.065,
and $\mu(T_1 )= 2.649(\pm 0.012) r^{1/4} + 12.832(\pm 0.034)$ with rms=0.069,
which are also plotted by the solid lines in the same figure.
It is found that the surface brightness profiles are well fit by a de Vaucouleurs  $r^{1/4}$ law.
% for $1.2 <r\arcmin_{maj}<9$.
The effective radius and standard radius for $C$-band (where $\mu(C)=25.0$ mag arcsec$^{-2}$) 
are derived to be 97 arcsec and 242 arcsec, %1.50 arcmin and 4.03 arcmin, 
respectively, corresponding to linear sizes of 8.15 kpc and 20.3 kpc. %7.56 kpc and 20.3 kpc. 
%These values are similar to previous estimats?????????????????%The eccentricity is 0.216.

%The radial profile of the $(C-T_1)$ color in the inner region of M60 shows clearly a negative radial gradient, 
%in that the color reddens as the galactocentric radius decreases. 
The $(C-T_1)$ color gets bluer in the inner region ($r\leq150\arcsec$) as the galactocentric radius increases, and
remains almost constant in the outer region. 
This trend for the inner region is consistent with the radial variation  of 
$(U-R)$ and $(B-R)$ colors given by \citet{pel90} as seen in Figure 10.

%($150\arcsec<r<500\arcsec$).
%The $(C-T_1)$ color gets much bluer again
%at $r>500\arcsec$, but this is due to the large photometric errors.
The color trend at $3\arcsec.2<r<151\arcsec.4$ can be fit by the log-linear
relation $(C-T_1 )=-0.162(\pm 0.011) {\rm \log} r + 2.160(\pm0.032)$ with rms=0.065.
The ellipticity increases rapidly from 0.1 to 0.2 in the central region, and stays
at an almost contant value in the outer region. 
The values of the ellipticity at the effective radius and standard radius
are derived to be, respectively, 0.21 and 0.22.
The PA increases from 90 deg to 100 deg in the central region, and changes
slowly in the outer region. The values of PA at the effective radius and standard radius
are, 106 and 110 deg, respectively.

\subsection{Spatial Distribution of the Globular Clusters}

We have investigated the spatial structure of the globular cluster system in M60.
Figure 11 displays the spatial distribution of the bright globular clusters
with $19<T_1<23$ mag (all globular clusters, BGCs and RGCs) 
as well as foreground stars found in the KPNO images.
We also plotted the very blue bright clusters (YC) with $(C-T_1 ) <0.5$ and $19<T_1<23$ mag.
We selected the red point sources with $(C-T_1)>2.5 $ and $T_1<23$ mag
as foreground stars for comparison. The spatial distribution of these
objects is seen to be uniform in Figure 11(b), showing indeed that they do not belong
to M60. 
We also created number density maps of the globular clusters,
which were smoothed and displayed in Figure 12. 
In Figure 12(b) we also display a grayscale map of the short exposure $T_1$ image of M60 that
shows the distribution of stellar light, which is to be compared  with the globular clusters.

Several notable features are seen in Figures 11 and 12.
First, the spatial distribution of all globular clusters is roughly circular, and it
 shows a strong central concentration.
Second, the spatial distribution of the BGCs is roughly circular and it  extends farther than that of the RGCs.
Third, the spatial distribution of the RGCs is somewhat elongated along the east-west direction, which
is consistent with the position angle of the halo of M60.
We derive e(ellipticity)= $0.05\pm0.02$, $0.03\pm0.02$, and $0.09\pm0.04$, respectively,
 for all globular clusters, BGCs and RGCs at $1\arcmin.5 <r< 7\arcmin$,
using the dispersion ellipse method of \citet{tru53}. 
Fourth, very blue, presumably young, bright clusters are located mostly within one arcmin from the center of NGC 4647 (inside the small circle at the position of NGC 4647 in Figure 11).

The Chanda X-ray image of M60 (in Fig. 1 of \citet{hum06}) shows that the X-ray emission of M60 is almost circular. Therefore the global structure of M60 seen in the X-ray image is closer to the spatial distribution of the BGCs rather than to that of the RGCs. This indicates that both the BGCs and the X-ray emitting hot gas may trace the dark mass of their host galaxies.

\subsection{Surface Number Density Profiles of the Globular Clusters}

We have derived the radial profiles of the surface number density of bright globular clusters with $19<T_1<23$ mag, using the KPNO data for the outer region at $r>1\arcmin.5$ and the HST data for the central region at $r<1\arcmin.5$. 
First we checked the radial profiles of mean counts per unit area for the objects with the same range of magnitude and color as the globular clusters in the KPNO images, finding that they get almost flat at $r>8\arcmin$. 
Therefore we derived the background levels from the mean
surface number density of these objects at $9\arcmin-10\arcmin$:
$1.988\pm 0.363$ per square arcmin for all  globular clusters, $1.790\pm 0.344$ per square arcmin for the BGCs, and  $0.199\pm 0.115$ per square arcmin for the RGCs.
Then we subtracted these background values from the original number counts
to produce the radial profiles of 
the net surface number density of globular clusters, which are listed in Table 4.
We derived the surface number density profiles in the core region using
the HST data with the same criteria as those for the KPNO data. 
We estimated the background contribution for the HST data 
using the result derived from the KPNO data, finding that it is negligible.

Figure 13 displays the radial profiles of the surface number density of globular clusters
in comparison with the $C$-band surface brightness profile of M60.
Several features are of note in Figure 13.
First, the surface number density profile of all globular clusters extends farther than the surface brightness profile of the galaxy halo. 
Second, the surface number density profile of the RGCs agrees approximately with the surface brightness profile of the galaxy halo, although the halo is steeper.
Third, the surface number density profile of the BGCs extends farther than that of the RGCs. 
Fourth,  the surface number density profiles of the globular clusters in the outer parts at $0\arcmin.5<r<7\arcmin$ are fit approximately both   by the de Vaucouleurs  $r^{1/4}$ law: 
$\log \sigma = -1.974(\pm0.047) r^{1/4} + 3.571(\pm 0.051) $ for all globular clusters, %at $0.5\arcmin <r< 7 \arcmin$,
$\log \sigma = -1.717(\pm0.068) r^{1/4} + 2.984(\pm 0.074 ) $ for the BGCs, and
$\log \sigma = -2.262 (\pm 0.050) r^{1/4} + 3.580(\pm 0.052) $ for the RGCs;
and by a power law:
$\log \sigma = -1.285(\pm 0.034) \log r + 1.563(\pm 0.011) $ for all globular clusters,
$\log \sigma = -1.128 (\pm 0.040) \log r + 1.237(\pm 0.014) $ for the BGCs,
and $\log \sigma = -1.452(\pm 0.050) \log r + 1.281(\pm 0.016) $ for the RGCs.
Effective radii are $8\arcmin.09$, $14\arcmin.14$, and $4\arcmin.70$ for all globular clusters,
BGCs and RGCs, respectively.
Thus the surface number density profile of the RGCs is much steeper than that of the BGCs.

Finally, the surface number density profiles of the globular clusters are flat in the central region at $r<0\arcmin.5$.
This result is affected little by the incompleteness of 
our photometry, because the completeness is higher than
95 \% for $V<23.5$ mag for all the radial range of the HST data.  
This flattening indicates the globular cluster system is dynamically relaxed in the central region in the sense that the radial profile of the central region can be fit by the King model \citep{kin62}.
Figure 14 shows the results derived from fitting the data for $r<2\arcmin$ with a King model.
From error-weighted fitting we derive core radius 
$r_c=0.84\arcmin$ and concentration parameter $c=3.91$ for all globular clusters, $r_c=1.06\arcmin$ and $c=3.92$ for the BGCs, and
$r_c=0.71\arcmin$ and $c=3.57$ for the RGCs.
From equal-weighted fitting we derive similar results:
$r_c=0.73\arcmin$ and $c=2.83$ for all globular clusters, 
$r_c=0.86\arcmin$ and $c=2.65$ for the BGCs, and
$r_c=0.66\arcmin$ and $c=2.98$ for the RGCs.
However, it is not easy to determine reliably the values of both parameters.
If we adopt a fixed value, $c=2.5$, as done for the case of M87 by \citet{kun99},
we derive $r_c=0.85\arcmin$ for all globular clusters, 
$r_c=1.06\arcmin$ for the BGCs, and $r_c=0.72\arcmin$ for the RGCs.
Thus the core radius for the RGCs is much smaller than that for the BGCs.
It is also noted that the surface number density profile of the RGCs at the outer area is reasonably fit by the King model, while that of the BGCs at $r>3\arcmin$ shows a clear excess over the King model. 
The central flattening is also seen in the  case of M49 \citep{lee00} and M87 \citep{kun99} both of which were also based on the HST/WFPC2 data.
The flattening for the globular clusters can be explained by the intrinsic property of the globular cluster formation epoch \citep{har98} or by the effect of the globular cluster destruction during the dynamical evolution \citep{cot98}. \citet{kun99} concluded there is little evidence
supporting the latter in the case of M87, noting that the luminosity function of the globular clusters in M87 does not vary spatially within about $1\arcmin.5$ from the center of M87.
Therefore the flattening for the M60 globular clusters may be the relic of the formation epoch
of the globular clusters.
%which is much larger than the core radius of the  

\subsection{Radial Variation of Mean Magnitudes and Colors}

Figure 15 displays the colors of the globular clusters with $19<T_1<23$ 
versus galactocentric distance derived from the KPNO data only. 
Mean and median colors of the globular clusters in each radial bin are represented 
by the larger symbols.
The mean color of the entire sample of globular clusters shows a clear
radial gradient, while those of the BGC and RGC
show little, if any, radial gradients. 
Linear least squares fitting for the range of  $1\arcmin <r< 9\arcmin$ yields
$(C-T_1)=-0.033(\pm 0.003) r + 1.718$ with rms=0.020 for all globular clusters,
$(C-T_1)=-0.015(\pm 0.003) r + 1.457$ with rms=0.012 for the BGCs, and
$(C-T_1)=-0.008(\pm 0.002) r + 1.984$ with rms=0.009 for the RGCs.
The samples at $r<1\arcmin$ in the KPNO data suffer from incompleteness so that
they were not included for fitting.
The mean color of the stellar halo is much closer to that of the RGCs than to that of the BGCs for $r<8\arcmin$, but is 0.2 to 0.3 mag bluer than that of the RGC.
This is in contrast to the case of M49 where the color profiles of the RGCs agree well with that of the stellar halo \citep{lee98}. 
The mean color for the RGCs depends on the color range adopted for taking a mean so that
the color difference between the halo and the RGCs is not considered to be significant.

Figure 16 displays the $C$ and $T_1$  magnitudes of the globular clusters with
 $19<T_1 <23$ mag versus galactocentric distance from the KPNO data only.
Mean magnitudes of the globular clusters in each radial bin are represented by the larger symbols.
The mean magnitudes of the globular clusters with $r>1\arcmin$ vary little
as a function of galactocentric distance. 
The upper (bright-side) envelope of the globular cluster magnitude distribution
also does not show any
clear systematic radial gradient.
Linear least squares fitting for the range of $1\arcmin <r< 9\arcmin$ yields
$T_1=0.007(\pm 0.008) r + 21.945$ with rms=0.053 for all globular clusters,
$T_1=0.012(\pm 0.017) r + 21.881$ with rms=0.058 for the BGCs, and
$T_1=0.033(\pm 0.013) r + 21.866$ with rms=0.050 for the RGCs.
The samples at $r<1\arcmin$ suffer from incompleteness problem so that
they were not included for fitting.
Thus we find little or no
dependence of the mean magnitude of the globular clusters on
galactocentric distance.

\subsection{Luminosity Function and Specific Frequency}

Figure 17 displays the luminosity function of the globular clusters in M60.
We plot the luminosity function of the globular clusters at $1\arcmin.5<r<8\arcmin$
derived from the KPNO photometry, 
the luminosity function of the background objects with the same color
range as that of the globular clusters
at $9\arcmin<r<10\arcmin$, and the net luminosity function for $1\arcmin.5<r<8\arcmin$ 
that was derived after
subtracting the background luminosity function from the luminosity function 
of the globular clusters. 
The luminosity functions were corrected for incompleteness using the completeness
values derived in Section 2.
We also plot  the luminosity function of the globular clusters at $r<1\arcmin.5$
derived from the HST photometry. 
The HST luminosity function of the globular clusters at $r<1\arcmin.5$
was derived by counting the globular clusters in a half circle of radius $r<1\arcmin.5$ and
doubling the resulting counts.
The KPNO luminosity function is similar to that of the HST luminosity function
for the bright magnitude range $T_1<23.5$, showing a peak at 
$T_1 \approx 23$. 
For $T_1>23.5$, the KPNO luminosity function falls below the HST luminosity function,
indicating that the KPNO photometry gets more incomplete in the faint magnitude range.
%Estimating N(total) and SN to compare  with F04, Ntotal=3653 for R<10'Sn=4
%             3700+/-900  (cf 5100+/-1100 Ashman \& Zepf 98)
\citet{lar01} derived the peak magnitude of the luminosity function of  the globular clusters in the central region of M60 from the HST/WFPC2 data: $V({\rm peak})=23.58\pm0.08$, which corresponds to $T_1 = 23.38$  
adopting the color $(C-T_1 ) =1.7$ in the equation given in the previous section, and the
foreground reddening $A_V=0.088$.
This value is consistent with the peak magnitude seen in the KPNO luminosity function.

We derived an estimate of the total number of globular clusters by 
(a) counting the brighter half of the luminosity function which is almost complete and 
(b) doubling it after subtracting background contribution, 
assuming that the luminosity function of the globular
clusters is symmetric around the peak magnitude.
The number of globular clusters brighter than the peak magnitude ($19<T_1<23.383$) at
$r<8\arcmin$ is    2048: 366 at $r<1\arcmin.5$ from the HST images, and 1682 at $1\arcmin.5<r<8\arcmin$ from the KPNO images. 
%1878: 254 at $r<1\arcmin.5$ from the HST images, and 1624 at $1\arcmin.5<r<8\arcmin$ from the KPNO images. 
We then derived the number of point sources with the same
color as the globular clusters in the background region at $9\arcmin<r<10\arcmin$ -- 40. %151.
If this number is scaled according to the area ratio of the regions at $r<8\arcmin$ 
and at $9\arcmin<r<10\arcmin$, it becomes 475. %151.
Part of these objects may be globular clusters, but we do not know what fraction of them.
So we arbitrarily assume a half of these as globular clusters
and assign the same value as an error. %an error, 237.5. %75.5.
Then the total number of the globular clusters is estimated to be $2(2048-237.5)=3621$.
%$2(1878-151)=3454$.
We take $N=3600\pm500$ %$N=3500\pm300$ 
as the total number of globular clusters in M60.
This estimate is very similar to the value derived by \citet{for04}, $3700\pm900$.
From this value and the absolute magnitude of M60 ($M_V = -22.44$) 
we derive the specific frequency to be $S_N = 3.8\pm 0.4$, which is similar to the value
given by \citet{for04}, $S_N\sim4.1\pm1.0$, and substantially smaller 
 than the value given by \citet{ash98},
$S_N\sim6.7$, based on inferior data. Our value, although on the small side for
a cluster gE, is not atypical.
%3600/10^{(22.44-15)/2.5}=3600/946.24=4.9
%3500/10^{(22.13-15)/2.5}=3500/711=4.9
%$S_N\sim4.1\pm1.0$, which is lower than the value $S_N\sim6.7$

\section{Discussion} 

\subsection{Comparison with Previous Studies} 

There is only one previous photometric study of the globular clusters of M60 covering
a relatively large field, that of \citet{for04} who covered about 90 square arcmin.
\citet{for04} showed that the color distribution of 995 bright globular clusters
with $20 < i < 23.6$ and $0.5<(g-i)<1.5$ in M60 is bimodal, with peaks at
$(g-i)=0.865\pm0.005$ and  $1.167\pm0.004$.
These peak colors correspond to %$(C-T_1 )=((g-i)-0.119)/0.527$
$(C-T_1 )=1.416$ and 1.989, respectively, using the transformation equation between
$(g-i)$ and $(C-T_1 )$ given in Section 2.5.
These values are similar to those derived in this study, $(C-T_1 )=1.37$ and 1.87.
Note that the color range between these two populations is almost twice as 
large in $(C-T_1)$ as $(g-i)$, illustrating the former's superior metallicity
sensitivity \citep{gei96b}. %(Geisler et al. 1996).

\citet{for04} also presented the surface number density profiles of the globular clusters
for the range of galactocentric distance $1\arcmin.3 - 3\arcmin.8$,
showing that the RGCs with $0.5<(g-i)<0.9$ have a similar surface density
distribution to that of stellar light, which is steeper than that
of the BGCs with  $1.1<(g-i)<1.5$. 
They found that the surface number density profiles of the BGCs, RGCs and all globular clusters are fit well
by power laws with slopes of $-1.04\pm0.09$, $-1.73\pm0.06$ and $-1.3\pm0.05$.
Their slopes for the BGCs and all globular clusters are similar to our estimates,
$-1.128\pm0.040$ and $-1.285\pm0.034$, but their slope of the RGCs is somewhat
steeper than ours, $-1.452\pm0.050$.

\citet{for04} presented also the radial variation of the mean 
colors of the globular clusters
for the range of galactocentric distance $0\arcmin.4 - 6\arcmin.3$,
showing that both the BGCs and RGCs show little radial color gradient,
while the color of all globular clusters gets bluer with increasing radius.
Our results on the radial gradient of colors shown in Figure 15 
are consistent with these results.

\subsection{Metallicity Distribution}

Colors of old globular clusters with similar  ages are determined primarily by the metallicity of the globular clusters.
There are several studies for deriving transformation relations between colors and [Fe/H]
for globular clusters, which were based on the integrated photometry of 
the Galactic globular clusters, %and some of which were based on 
the combined data of globular clusters in a few galaxies,
or stellar population synthesis models \citep{gei90,har02,coh03,pen06,yoo06}.

However, there are only a few studies on the transformation relation between 
$(C-T_1)_0$ and [Fe/H] for globular clusters in the literature. 
\citet{gei90} presented a linear relation derived from the data for the Galactic globular clusters: [Fe/H] $ = 2.35(C-T)_0 -4.39$;
\citet{har02} derived a quadratic relation based on the data for the Galactic globular clusters: [Fe/H]$ = -6.037[1-0.82 (C-T_1 )_0 + 0.162{(C-T_1 )_0}^2]$; and
\citet{coh03} derived another quadratic relation based
on the combined data including the globular clusters in M49 and M87 as well as the Galactic globular clusters:
[Fe/H]$= -0.75(C-T_1 )_0^2 +4.438 (C-T_1 )_0 -5.64$.
However, the sequence for M87 globular clusters is not consistent with those of M49 globular clusters and Galactic globular clusters in the [Fe/H] versus $(C-T_1 )_0$ diagram
\citep{coh03}, while the sequences for the latter two are consistent.
The [Fe/H] data for 38 globular clusters in M60 presented by \citet{pie06} show
too large a scatter to be used for this purpose.

We derived new relations using the combined data for the globular clusters
 in our Galaxy and M49.
The [Fe/H] data are from \citet{har96} for Galactic globular clusters,
and \citet{coh03} for M49 globular clusters.
The $(C-T_1  )$ data are from \citet{har77} for Galactic globular clusters, and
from \citet{gei96b, lee98} for M49 globular clusters.
We updated the $(C-T_1 )$ data for the Galactic globular clusters 
by adding our unpublished data for two globular clusters, NGC 6624 and NGC 6316,
 $(C-T_1 )=2.75\pm0.02$, and  $2.096\pm0.02$, respectively.
We applied the reddening correction using the values given in \citet{har96}.

Figure 18 displays a [Fe/H] versus $(C-T_1 )_0 $ diagram for these data.
The sequence for the M49 globular clusters (open squares) 
matches
well that for the Galactic globular clusters, and covers as well 
the high [Fe/H] range where there are only a few data for Galactic globular clusters. 
The relation between [Fe/H] and $(C-T_1 )_0 $ appears approximately linear 
over most of the  range, 
with some possible    curvature      at both ends, especially at the metal-poor
end.
We tried to fit the combined data of our Galaxy and M49, after removing outliers,
using various equations, and found 
that a 3rd order polynomial and double linear relations give the best fits: 
[Fe/H]$=1.387 (C-T_1 )_0^3 -6.698 (C-T_1 )_0^2 + 12.609 (C-T_1 )_0 -9.379$ 
with reduced $\chi^2 =  2.529$,  and  
[Fe/H] $= (2.359\pm0.051) ((C-T_1 )_0 -1.46) -0.92 $  for $(C-T_1 )_0 \le 1.46$,
 $ = (1.951 \pm 0.044) ((C-T_1 )_0 -1.46) -0.92$ for  $(C-T_1 )_0 >1.46$ % with rms=0.121 and
 with reduced $\chi^2 =  5.090$. 
These relations are similar in general to those given in the literature, 
showing some difference only around the low and high metallicity ends.
 But note that the \citet{coh03}  curve is 
significantly displaced from the other relations at virtually all 
metallicities.

We derived [Fe/H] for M60 globular clusters from 
$(C-T_1)_0$ colors using our relations and previous relations \citep{gei90,har02,coh03}, 
and displayed  them in Figure 19.
We plotted also the [Fe/H] distribution from \citet{har96} %Harris (1996)
for the Galactic globular clusters for comparison
in the same figure.
Figure 19 shows several notable features as follows.
First,  the [Fe/H] distributions are very broad, covering from [Fe/H]$\approx -2.4$
to much higher than the solar value,  reaching [Fe/H]$\approx 0.8 $ and much higher than in
the Galaxy.
Second, all the [Fe/H] distributions derived from various transformation relations show
a dominant peak at [Fe/H]$\approx -1.2$ dex. 
Third, none of the [Fe/H] distributions derived from various transformation relations
look symmetric.
Most of them show clearly a weaker component  at [Fe/H]$\approx -0.2$ dex in addition to the dominant peak at [Fe/H]$\approx -1.2$ dex, 
although the \citet{har02} calibration shows a dominant metal-rich peak.
Thus they are all bimodal in metallicity. 
A typical photometric error of $\sigma(C-T_1 )=0.1$
leads to an error $\sigma$[Fe/H]$=0.20 \sim 0.24$ when using the two transformation relations derived in this study. 

KMM tests of the data based on the 3rd order polynomial transformation show that the probability
that the [Fe/H] distribution is  bimodal over unimodal is  higher than 99.9 \%. 
%can be fit better by two Gaussians rather than by a single Gaussian.
We determined the parameters for best-fit double Gaussian curves to the metallicity
distribution data for 1236 globular clusters, using
a maximum-likelihood method through the KMM mixture modeling routine \citep{ash94}:
%KMM double Gaussian fits to the data for 1,236 globular clusters yield  
a metal-poor component with center at [Fe/H]$ = -1.18$, $\sigma=0.42$, and N=779,
and a metal-rich component with center at [Fe/H]$ = -0.27$, $\sigma=0.59$, and N=467 
for  the 3rd order polynomial transformation. Similar results are obtained for 
the double linear transformation: 
a metal-poor component with center at [Fe/H]$ = -1.22$, $\sigma=0.39$, and N=747, 
and a metal-rich component with center at [Fe/H]$ = -0.18$, $\sigma=0.44$, and N=489. 
Fourth, the metallicities of both peaks in M60 are 0.3 -- 0.4 dex larger, respectively,
than those for Galactic globular clusters at [Fe/H]$=-1.5$ and $-0.6$.

We compared [Fe/H] derived using the double linear relation in this study 
with the estimates
given by \citet{pie06} for 38 common globular clusters, as shown in Figure 20.
\citet{pie06} derived [Fe/H] of 38 globular clusters using two methods from spectra:
one using the simple stellar population (SSP) models and the other using the Brodie \& Huchra (BH) method \citep{bro90}.
Figure 20 displays SSP and BH metallicities given by \citet{pie06} 
versus  [Fe/H] derived in this study.
It is seen that two  show good correlation but with some scatter.
In addition, \citet{pie06}'s [Fe/H] is on average 0.3 to 0.4 dex lower than
our [Fe/H]. Linear fits to the data yield:
SSP [Fe/H](Pierce et al.) = 1.120[Fe/H](this study) --0.232 with rms=0.408, and
BH [Fe/H](Pierce et al.) = 0.805[Fe/H](this study) --0.558 with rms=0.315.
The cause for this difference is not known.
%Given the respective errors in these methods
%the scatter is as expected?????????????.

\subsection{Comparison with M49 (NGC 4472)}

There are several gEs in Virgo and they are an excellent pool for studying the 
properties of the globular clusters in gE, because they belong to the same galaxy cluster,
being located at the similar distance from us. 
M49 (NGC 4472)  is the brightest gE in Virgo, located about 4 deg from the  Virgo center.
It is a prototypical example of a galaxy showing the bimodal color distribution
of the globular clusters in gEs.
Detailed analysis of the Washingtong $CT_1$ photometry of the globular cluster system in M49 based on similar data was given by \citet{gei96b} and \citet{lee98}.
Here we compared in detail the properties of the globular cluster systems in M60 and M49,
using very similar data and techniques. We have also limited ourselves to
almost identical globular cluster sample definitions: $r=1-7\arcmin$ and $T_1=19-23$. %, and $(C-T_1)=1-2.4$.
The color boundary for M49  ($(C-T_1)=0.9-2.3$) is 0.1 bluer than that for M60.

Figure 21 displays the color distribution of the bright globular clusters in M60
and M49. It shows that the color distributions of the bright globular clusters in
both galaxies are similarly bimodal.
We determined the parameters for best-fit double Gaussian curves to the color
distribution data for M49, using
a maximum-likelihood method through the KMM mixture modeling routine: % \citep{ash94}:
a primary component with center at $(C-T_1)=1.30$, $\sigma=0.13$, and N=792, 
%width(FWHM) of 0.38, and
and  a secondary component with center at $(C-T_1)=1.79$, $\sigma=0.21$, and N=818.
One difference between M60 and M49 is 
that the peak colors for M60 ($(C-T_1)=1.37$ and 1.87) 
are slightly redder than those for M49 ($(C-T_1)=1.30$ and 1.79), which is consistent
with the finding by \citet{cou91} based on $BV$ photometry.
The difference in the foreground reddening  between M60 ($E(B-V)=0.026$, $E(C-T_1)=0.051$) and M49 ($E(B-V)=0.022$, $E(C-T_1)=0.043$) is negligible \citep{sch98} .
Therefore the colors of the globular clusters in M60 are considered to be on average
intrinsically about 0.1 mag redder than those in M49. 
If the globular clusters are of similar old ages, this implies that the M60 globular clusters are on average
about 0.2 dex more metal-rich than their M49 counterparts.

Recently \citet{str07} found from the re-analysis of the spectroscopic data for 47 bright
globular clusters in M49 \citep{coh03} that the metallicity distribution of these globular clusters
is bimodal with two peaks at [m/H]$\sim -1.1$ and 0.0.
The peak metallicity for the metal-poor globular clusters in M49 is $\sim 0.4$ dex 
higher than that for the metal-poor globular clusters in our Galaxy, and 
the peak metallicity for the metal-rich globular clusters in M49 is $\sim 0.5$ dex 
higher than that for the metal-rich globular clusters in our Galaxy.
These results are similar to those derived from the Washington photometry of the globular clusters in M49 and M60 in this study. 

The spatial distribution of the globular clusters in both galaxies share common features:
(a) the RGCs are more centrally concentrated than the BGCs, 
(b) the globular cluster system is more extended than the stellar halo, and 
(c) the elongation of the RGC system is consistent with
that of the stellar halo, while that of the BGC system is approximately
circular,
(d) the color gradient in the overall globular cluster systems are similar, as are the relative
lack of color gradients in the individual RGC and BGC populations.

Recent studies based on the analysis of HST ACS/WFC data for the globular clusters
in elliptical galaxies found that the brighter the bright BGCs are, the redder their colors get,
which has been referred to as the `blue tilt' \citep{har06,str06, mie06}.
Interestingly, previous studies have found significant blue tilts in M87 and M60, 
but no evidence in M49 \citep{str06, mie06}. All these results were based on 
the HST observation of small fields mostly covering the central regions of the galaxies.
\citet{mie06} found also that the blue tilts are seen in both inner regions (at $r<65\arcsec$) and ``outer" regions (at $r>65\arcsec$) 
of bright galaxies, and that the slope for the
globular clusters in the inner region is two-to-three times steeper in
$d(g-z)/dMz$ than that in the outer region.

Figure 22 displays the color-magnitude diagrams of M60 and M49 derived 
from the KPNO images for the outer regions at $1\arcmin.5 <r <7\arcmin$.   
We used the data for M49  from \citet{gei96b} and \citet{lee98}. 
We determined the parameters for the best-fit double Gaussian curves describing the color
distribution for given magnitude range, using
a maximum-likelihood method through the KMM mixture modeling routine, and plotted  
the center values of the Gaussian components in Figure 22.

In Figure 22 the blue tilt  is clearly seen for the bright BGCs in M60, 
while it is not clearly seen for the bright RGCs. 
Only the bright BGCs with $T_1 \leq 22$ in M60 show this blue tilt, 
which  is consistent
with the finding for the brightest cluster galaxies by \citet{har06} that the bright BGCs
with $M_I \leq -9.5$ show the blue tilt, while the faint BGCs $M_I > -9.5$ do not.
The blue tilt is also seen for the bright BGCs with $T_1<22$ mag in M49, 
but it is much weaker than that for M60. 
The color change for the BGCs with $20<T_1 <22$ is about 0.2 mag for M60, while it is about 0.1 mag for M49. 
However, the brightest bin for M60 is not separated clearly into two groups so that the difference in color change between M60 and M49 may not be significant.
This is in contrast with the case of the inner region of M49 where no blue tilt was found in the HST ACS images \citep{str06,mie06}.
Interestingly the RGCs in M49 also show a similar tilt like the BGCs, which is different  
from the typical CMD. Therefore it is needed to investigate this problem with better data.

%We derived the slopes of the blue tilt using  linear least-squares fitting to the Gaussian center values of the color for the bright BGCs with $20.5<T_1<22$ ($-10.8 < M_{T1}<-9.3$),
%listing them in Table 5. 
%We did not include the brightest bin for fitting, because the brightest bin has a small number of data.
%, and the faintest bin includes the data with relatively larger photometric errors. 
%Two features are noted in Table 5.
%First, the slopes of the blue tilt do not show little, if any, systematic variation depending on the galactocentric distance  in both galaxies. 
%The only exception is that for the inner region in M49, where
%the blue tilt is not seen, and in fact a 'red tilt' appears.
%Second, the slopes of the blue tilt for the range of 1.5 to 7 arcmin are similar
%between both galaxies: %$-0.062\pm0.014$ for M60 and $-0.077\pm0.012$ for M49.
%$-0.119\pm0.023$ for M60 and $-0.098\pm0.021$ for M49.

Several mechanisms were discussed for the origin of the blue tilt \citep{har06,str06, mie06, bek07}: 
accretion of globular clusters from low-mass galaxies, the capture of field stars
by individual globular clusters,  self-enrichment in globular clusters, 
contamination effect (due to super-star clusters, stripped nuclei, and 
ultra compact dwarf galaxies (UCDs)), 
and stochastic effects.
After examining these, \citet{mie06} concluded that self-enrichment and field star capture
are the most promising to explain the observational results for the blue tilt derived
from the HST ACS data for a sample of 79 early-type galaxies.
 
We discuss these two mechanisms in relation with the observational results for M60 and M49.
If the blue tilt is mainly due to field star capture, it is expected that there is a radial
variation of the blue tilt in the sense that it is steeper in the inner region. 
This is because there is a negative radial color gradient for halo stars at $r<2\arcmin.5$ 
as seen in the surface color profiles, 
and because more massive globular clusters can collect more
field stars that are redder in the inner region.
However it is not possible to investigate this trend reliably with the current data for M60 and M49.
%the current data for M60 and M49. do not allow to investigate this trend further.
%So the possibility of field star capture
%is not consistent with the observational results for M60 and M49.

On the other hand, the degree of self-enrichment is affected mainly by the mass of the stellar systems.
So the self-enrichment effect of the globular clusters is expected to depend on the mass of the globular clusters, but little on the galactocentric distance. This is consistent with
the observational results for the outer regions of M60 and M49.  
However, the existence of self-enrichment is seen in few, if any, typical globular clusters in our Galaxy and is a controversial subject (\citet{tho02} and references therein). 
It is also difficult to explain the absence of the blue tilt in the inner region of M49
with self-enrichment. % without invoking a merging scenario depending on galaxies.
Further studies of globular clusters covering the full spatial extent of the
globular cluster systems of gEs are needed to understand this problem.  

\subsection{Implications for the Formation of Globular Clusters}

Key observational facts for globular clusters in M60 to be explained by any formation scenario 
can be summarized as follows.
1) The color distribution of the globular clusters is bimodal.
2) This indicates that the range of metallicity of the globular clusters is very large
from [Fe/H]$< -2.0$ to $\sim 0.5$, and that the metallicity distribution is
also bimodal, consisting        of 
a strong metal poor component with peak [Fe/H]=$-1.2$ 
and a weaker metal rich component with peak [Fe/H]$\approx -0.2$.
3) The spatial distribution -- density, ellipticity and position angle --
 of the RGCs is similar to that of halo stars, while the
spatial distribution of the BGCs is much more extended and circular than that of the stellar halo.
4) The bright BGCs show a blue tilt, while the bright RGCs do not.

Several scenarios to explain the formation of globular clusters in gEs have been proposed: primordial formation \citep{pee68} including the biased globular cluster formation by \citet{wes93}, 
gaseous merger \citep{ash92}, in situ multiphase collapse model \citep{for97, kis98}, and tidal stripping/accretion \citep{cot98}. 
Each of the scenarios has pros and cons, but none of them by itself 
can explain all the observational results.
The current paradigm in the formation scenarios of the globular clusters addresses the
following three processes  \citep{lee03,kra05,bro06,har06}.

\begin{enumerate}

\item Formation of the first generation of globular clusters: The metal-poor globular clusters
were formed first among the stellar systems in the universe \citep{pee68}. 
The mass of their progenitors must
be much larger than the mass of the present globular clusters, possibly similar to that of dwarf
galaxies ($10^8-10^9$ $M_\odot$), to explain the blue tilt via self enrichment.
They were formed not only around massive galaxies, but also in the field throughout the universe where the gas density was sufficiently high.
The latter could now be intracluster globular clusters. 
However, the fact that the metal-poor globular clusters
are found in and around the galaxies 
%and the fact that the intracluster globular clusters are found only in galaxy clusters 
implies that they were formed preferentially around the high density
peaks rather than in any random field in the universe, as proposed by \citet{wes93}.

\item Formation of the second generation of globular clusters: 
The metal rich globular clusters were formed together with the halo stars, both of which were formed %, via dissipative collapse during mergers,
in the interstellar medium that was  enriched rapidly after the formation of the first generation
of globular clusters. This enrichment time scale should be smaller than about two Gyr \citep{lee03}, and the enrichment
range should be large, from [Fe/H]$<-1.2$ to $\approx -0.2$,
a range similar to that between the metal-poor and metal-rich Galactic globular cluster
populations but at a significantly higher metallicity.
The enrichment process can be double-episodal or continuous, 
depending on the existence of two peaks or a single broad peak in the metallicity distribution 
\citep{bea02}. %(e.g. Beasley et al. 2000???).
The broad metallicity distribution and the high metallicity peak being
 much weaker than the low metallicity peak, if anything, indicate that the enrichment process might have been mainly
continuous rather than double-episodal. However, the existence of a minor high metallicity component
indicates that there is a separate component of globular clusters that were formed 
in another episode. 
The color gradient of these globular clusters is very similar to that of the halo stars.
These globular clusters in giant elliptical galaxies 
can be formed via gaseous merger of galaxies.
If they are formed via gaseous merger, they must have been formed during the early
phase of merging (within about two Gyr from the birth of their host galaxies).
Later merging can make the host galaxies grow mostly in mass, not in metallicity \citep{del06}.
It is expected that the spatial distribution of these metal-rich globular clusters and halo
stars is more centrally concentrated compared with that of the metal-poor globular
clusters.
The spatial distribution of the metal-rich globular clusters can be similar to or 
slightly more extended than that of the halo stars, depending on the formation time scale
and dissipation effect of stars and globular clusters (see also \citet{tam06b}).

\item Tidal capture/Accretion of the existing globular clusters:
 Massive galaxies can collect globular clusters in the nearby surroundings that may include both
lower mass galaxies or the field. 
Globular clusters accreted after about one or two Gyr from the first star formation
can have any range of metallicity. However, most of the accreted globular clusters are
probably metal-poor rather than metal-rich, 
since low mass galaxies have generally more metal-poor globular clusters
\citep{mil06}, %(ref. B. Miller/Lotz work???), 
and they are more likely to be seen in the outer regions of massive galaxies.

\end{enumerate}

Observational results for the globular clusters in M60 show features of all three processes.
The metal-poor globular clusters are involved with the first and third processes, while
the metal-rich globular clusters are related with the second process.
Therefore, all three mechanisms may have played a role in the formation of
the M60 globular clusters.

%biased destruction of globular clusters vs biased formation????

\section{Summary and Conclusion}

We have presented Washington $CT_1$ photometry of the globular clusters
in M60 as well as that of the star clusters in NGC 4647, a companion spiral
galaxy, covering a $16\arcmin.4 \times 16\arcmin.4$ field. 
We have analyzed various photometric properties of the globular clusters in M60.
Primary results are summarized as follows.

\begin{enumerate}

 \item 
The color-magnitude diagram reveals a significant population 
of globular clusters in M60, and a large number of young luminous clusters
in NGC 4647.

\item 
The color distribution of the globular clusters in M60 is clearly bimodal,
with a blue peak at $(C-T_1)=1.37$, and a red peak at at $(C-T_1)=1.87$.

\item  We derived two new transformation relations between the $(C-T_1 )_0$ color and [Fe/H]
using the data for the globular clusters in our Galaxy and M49:
a 3rd order polynomial relation, 
[Fe/H]$=1.387 (C-T_1 )_0^3 -6.698 (C-T_1 )_0^2 + 12.609 (C-T_1 )_0 -9.379$, %with reduced $\chi^2 =  2.529$,  
and  a double linear relation, 
[Fe/H] $= 2.359 ((C-T_1 )_0 -1.46) -0.92 $  for $(C-T_1 )_0 \le 1.46$, and
          $ = 1.951 ((C-T_1 )_0 -1.46) -0.92$ for  $(C-T_1 )_0 >1.46$.
% with rms=0.121 and reduced $\chi^2 =  5.090$. 
Using these relations we derived the metallicity distribution of the globular clusters in M60,
which is bimodal: 
a dominant metal-poor component with center at [Fe/H]$ \approx -1.2$,
% $\sigma=0.40$, and N=792, %765,
and a weaker metal-rich component with center at [Fe/H]$ \approx -0.2$.
%, $\sigma=0.42$, and N=444. 

\item 
The radial number density profile of the globular clusters is more extended than that of the stellar halo.  
The radial profiles for the outer region at $1\arcmin<r<7\arcmin$ are fit 
approximately  equally well by the deVaucouleurs law and a  power law.
The radial  number density profile of the BGC is more extended that that of the RGCs.
The radial profiles for the central region show flattening, and the radial profiles for $r<2\arcmin$ is
well fit by the King model.
The core radii derived for the fix concentration value of $c=2.5$ are
I$r_c=0.85\arcmin$ for all globular clusters, 
$r_c=1.06\arcmin$ for the BGCs, and $r_c=0.72\arcmin$ for the RGCs.
Thus the core radius for the RGCs is much smaller than that for the BGCs.
\item
The surface number density maps of the globular clusters show
that the spatial distribution of the BGCs is roughly circular,
while that of the RGCs is elongated similarly to that of the 
stellar halo.

\item
The mean color of the bright BGCs gets redder as they get brighter
both in the inner and outer region of M60. This blue tilt is seen also in the outer region
%, but not in the inner region 
of M49, the brightest Virgo galaxy.

\item
We estimated the total number of the globular clusters in M60 to be $3600\pm500$,
and the specific frequency to be $S_N=3.8\pm0.4$.

\end{enumerate}

\acknowledgments
The authors are grateful to the anonymous referee for very detailed
and useful comments that improved the original manuscript significantly. 
This work was supported in part by a grant (R01-2007-000-20336-0) from the Basic Research Program of the Korea Science and Engineering Foundation.
D.G. gratefully acknowledges support from the Chilean 
{\sl Centro de Astrof\'\i sica} FONDAP No. 15010003.
%We are grateful to V. Barger, T. Han, and R. J. N. Phillips for
%doing the math in section~\ref{bozomath}.
%More information on the AASTeX macros package is available \\ at
%\url{http://www.aas.org/publications/aastex}.
%For technical support, please write to
%\email{aastex-help@aas.org}.

%% in parentheses, after the keyword, but they will not be verified.

{\it Facilities:} \facility{KPNO}, \facility{HST (WFPC2)}. %, \facility{CXO (ASIS)}.

\clearpage
\begin{deluxetable}{ccc}
%\tabletypesize{\scriptsize}
%\rotate
\tablecaption{Basic information of M60 (NGC 4649) \label{table1}}
\tablewidth{0pt}
\tablehead{
\colhead{Parameter} & \colhead{Values } & \colhead{References} }
\startdata
RA(2000), Dec(2000) & 12h43m39.66s,  +11$^\circ$ 33$\arcmin$ 9$\arcsec.4$ & 1 \\
Effective radius, $R_{eff}$    & 97 arcsec ($C$), 110 arcsec ($T_1$) & 2\\
Effective ellipticity, $e_{eff}$ & 0.21 ($C$,$T_1$ ) & 2 \\
P.A. ($R_{eff}$) & 106 deg ($C$), 105 deg  ($T_1$)  & 2 \\
Standard radius, $R_{25}$     & 242 arcsec ($C$)& 2\\
Standard ellipticity, $e_{25}$  & 0.224 ($C$) & 2 \\
P.A. ($R_{25}$) &  108 deg ($C$) & 2 \\
Total magnitudes              & $V^T=8.84\pm0.05$, $B^T=9.81\pm0.05$ & 3\\
X-ray luminosity   &  $Log(L_X) = 41.16$ [erg s$^{-1}$] & 4 \\
%$Log(L_x) = 41.12$ [erg s$^{-1}$] at d=15.92 Mpc & 4 \\
Systemic radial velocity, $v_p$                 & $1117\pm6$ km s$^{-1}$ & 1\\
Foreground reddening         & $E(B-V)=0.026$  & 5\\
Distance &  d=17.30 Mpc ($(m-M)_0=31.19\pm0.07$)  & 6 \\
\enddata
%% Text for table notes should follow after the \enddata but before
%% the \end{deluxetable}. Make sure there is at least one \tablenotemark
%% in the table for each \tablenotetext.
%\tablecomments{Table \ref{tbl-1} is published .}
%\tablenotetext{a}{Sample footnote t}
%\tablenotetext{b}{Another sample footnote for table~\ref{tbl-1}}
\tablerefs{(1) NASA Extragalactic Database; %, \citet{gon93};
 (2) This study;
(3) \citet{dev91}; (4) \citet{osu01}; (5) \citet{sch98}; (6) \citet{mei07}.}
%???need to include Dec???
%(25) Hobbs \etal 1991; (26) Olsen 1983.}
\end{deluxetable}

\clearpage

\begin{deluxetable}{cccccc}
%\tabletypesize{\scriptsize}
%\rotate
\tablecaption{Observation log \label{table2}}
\tablewidth{0pt}
\tablehead{
\colhead{Target} & \colhead{Filter} & \colhead{T(exp)} & \colhead{Airmass} & \colhead{Seeing} &
\colhead{Date(UT)}}
\startdata
M60 & C & 100 & 1.4 & $1\arcsec.43$ & 1997 Apr 9 \\
M60 & C & 1500 & 1.5 & $1\arcsec.47$ & 1997 Apr 9 \\
M60 & R & 60 & 1.3 & $1\arcsec.43$ & 1997 Apr 9 \\
M60 & R & 1000 & 1.4 & $1\arcsec.35$ & 1997 Apr 9 \\
M60 & C & $3 \times 1500$ & 1.2 & $1\arcsec.09-1\arcsec.24$ & 1997 Apr 10 \\
M60 & R & $2\times 1000$ & 1.4 & 1.34-$1\arcsec.44$ & 1997 Apr 10 \\
\enddata
%% Text for table notes should follow after the \enddata but before
%% the \end{deluxetable}. Make sure there is at least one \tablenotemark
%% in the table for each \tablenotetext.
%\tablecomments{Table \ref{tbl-1} is published .}
%\tablenotetext{a}{Sample footnote t}
%\tablenotetext{b}{Another sample footnote for table~\ref{tbl-1}}
%\tablerefs{(1) Barbuy, Spite, \& Spite 1985; (2) Bond 1980; (3) Carbon \etal 1987;
%(25) Hobbs \etal 1991; (26) Olsen 1983.}
\end{deluxetable}
%\clearpage

\begin{deluxetable}{c rrrr cccc}
%\tabletypesize{\scriptsize}  
%%%%%%% 
%%%%% maketable1.sm -> textab0.txt for good point sources with e(c-t1)<0.3. N=4497
%\rotate
\tablecaption{$CT_1 $ photometry of the measured point sources with $\sigma(C-T_1 ) <0.3$ in M60$^a$ \label{table3}}
\tablewidth{0pt}
\tablehead{
\colhead{ID} & \colhead{X[px]$^b$} & \colhead{Y[px]$^b$} & \colhead{RA[deg]} & \colhead{Dec[deg]} & \colhead{$T_1$} & \colhead{$\sigma$($T_1$)} &
\colhead{$(C-T_1)$} & \colhead{$\sigma$($C-T_1$)} }
\startdata
%0.89999 & 111 & 111 & 00 00 00.00 & 00 00 00.0 & 20.000 & 0.005 & 1.234 & 0.005 \\
   2  &   182.38 &  457.14 & 190.800629 &  11.479574 &  17.714 &   0.039  & 3.265 & 0.039 \\
   4  & 991.59  & 1164.33  & 190.908737 &  11.571802 &  17.898 &  0.033 &  2.386  &  0.038 \\
   5  & 665.44  & 1204.13  & 190.865173 &  11.577062 &  17.925 &  0.025 &  1.095 &   0.031 \\
   6  & 144.16 &  112.40  &  190.795517  & 11.434722 &  18.048  & 0.063 &  3.753  &  0.066 \\
   7  & 862.64 &  923.39  &  190.891418 & 11.540276 &  18.060  & 0.014 &  2.204  &  0.028 \\
%      8    1352.43     102.93   12.730435   11.432987   18.193    0.064    2.859    0.065 
%      9     461.29     572.10   12.722520   11.494430   18.268    0.029    2.292    0.030 
%     10    1804.00     436.51   12.734455   11.476486   18.359    0.029    3.965    0.036 
%     11     846.04     558.06   12.725945   11.492471   18.377    0.033    2.732    0.034 
%     12    1433.82    1051.79   12.731184   11.556989   18.378    0.026    2.507    0.028 
\enddata
\tablenotetext{a}{The complete version of this table is in the electronic edition 
of the Journal. The printed edition contains only a sample.}
\tablenotetext{b}{X and Y increases toward east and north, respectively. A pixel corresponds to 0.47 arcsec.}
%% Text for table notes should follow after the \enddata but before
%% the \end{deluxetable}. Make sure there is at least one \tablenotemark
%% in the table for each \tablenotetext.
%\tablecomments{Table \ref{tbl-1} is published .}
%\tablenotetext{a}{Sample footnote t}
%\tablenotetext{b}{Another sample footnote for table~\ref{tbl-1}}
%\tablerefs{(1) Barbuy, Spite, \& Spite 1985; (2) Bond 1980; (3) Carbon \etal 1987;
%(25) Hobbs \etal 1991; (26) Olsen 1983.}
\end{deluxetable}

%% If you use the table environment, please indicate horizontal rules using
%% Do not put multiple tabular environments within a single table.
%% The optional \label should appear inside the \caption command.

%\clearpage

\begin{deluxetable}{cccc} 
%\tabletypesize{\scriptsize}
%\rotate
\tablecaption{Number density profiles of globular clusters with $19<T_1<23$
in M60 \label{table4}} %% source: m60khdent23b.dat
\tablewidth{0pt}
\tablehead{
\colhead{r[arcmin]} & \colhead{$\sigma$(All GC)$^a$} & \colhead{$\sigma$(BGC)$^a$} & \colhead{$\sigma$(RGC)$^a$}}
\startdata
0.083 & $115.870\pm37.203$ & $45.474\pm23.529$ & $70.394\pm28.818$ \\
0.292 & $85.695 \pm 15.463$ & $ 33.950\pm 9.856 $ & $ 51.742\pm11.915 $ \\
0.542 & $ 78.312\pm12.663 $ & $34.455 \pm 8.495$ & $43.855 \pm 9.391$ \\
0.833 & $ 47.184\pm 6.602$ & $20.670 \pm 4.451$ & $26.512 \pm 4.876$ \\
1.167 & $ 29.771\pm 4.756$ & $ 14.189\pm3.363 $ & $ 15.580\pm 3.363$ \\
1.750 & $ 20.160\pm 2.082$ & $ 10.668\pm 1.556$ & $ 9.490\pm 1.383$ \\
2.250 & $ 14.561\pm 1.626$ & $ 6.988\pm1.178 $ & $ 7.571\pm 1.121$ \\
2.750 & $ 10.288\pm 1.258$ & $ 5.496\pm 0.964$ & $ 4.789\pm 0.808$ \\
3.250 & $ 9.890\pm 1.112$ & $ 5.945\pm 0.894$ & $ 3.942\pm 0.662$ \\
3.750 & $ 5.714\pm 0.799$ & $ 2.586\pm0.596$ & $3.126\pm0.532$ \\
4.250 & $ 5.339\pm 0.730$ & $ 3.434\pm 0.613$ & $1.903 \pm0.396 $ \\
4.750 & $ 5.059\pm 0.677$ & $ 3.241\pm 0.569$ & $1.817 \pm 0.367$ \\
5.250 & $ 3.921\pm 0.588$ & $ 2.780\pm 0.514$ & $ 1.140\pm 0.284$ \\
5.750 & $ 3.983\pm 0.565$ & $ 2.846\pm 0.495$ & $ 1.135\pm 0.271$ \\
6.250 & $ 1.990\pm 0.438$ & $ 1.367\pm 0.388$ & $ 0.620\pm 0.204$ \\
6.750 & $ 2.940\pm 0.472$ & $ 1.906\pm 0.406$ & $ 1.032\pm 0.241$ \\
7.250 & $ 2.086\pm 0.412$ & $ 1.620\pm 0.375$ & $ 0.465\pm 0.170$ \\
7.750 & $ 1.174\pm 0.355$ & $ 0.425\pm 0.293$ & $ 0.747\pm 0.201$ \\
8.250 & $ 1.040\pm 0.411$ & $ 0.632\pm 0.365$ & $ 0.405\pm 0.190$ \\
8.750 & $ 1.139\pm 0.493$ & $ 0.831\pm 0.449$ & $0.306 \pm 0.204$ \\
\enddata
\tablenotetext{a}{Number per square arcmin after background subraction.}
%% Text for table notes should follow after the \enddata but before
%% the \end{deluxetable}. Make sure there is at least one \tablenotemark
%% in the table for each \tablenotetext.
%\tablecomments{Table \ref{tbl-1} is published .}
%\tablenotetext{a}{Sample footnote t}
%\tablenotetext{b}{Another sample footnote for table~\ref{tbl-1}}
%\tablerefs{(1) Barbuy, Spite, \& Spite 1985; (2) Bond 1980; (3) Carbon \etal 1987;
%(25) Hobbs \etal 1991; (26) Olsen 1983.}
\end{deluxetable}

\clearpage

%% the file, uncomment the \plotone, \plottwo, and
%% If you need a layout that cannot be achieved with \plotone or
%% \plottwo, you can invoke the graphicx package directly with the
%% \includegraphics command or use \plotfiddle. For more information,
%% please see the tutorial on "Using Electronic Art with AASTeX" in the
%% documentation section at the AASTeX Web site,
%% http://www.journals.uchicago.edu/AAS/AASTeX.
%%
%% The examples below also include sample markup for submission of
%% supplemental electronic materials. As always, be sure to check
%% the instructions to authors for the journal you are submitting to
%% for specific submissions guidelines as they vary from

%% This example uses \plotone to include an EPS file scaled to
%% 80% of its natural size with \epsscale. Its caption
%% has been written to indicate that additional figure parts will be
%% available in the electronic journal.

%%%%%%%%%%%%%%%%%%%%%%%%%%%%%%%%%%%%%%%%%%%%%%\input{figcap4}       %Apr 23, 2007
%May 31, Apr 23, 2007 for M60

%% This example uses \plotone to include an EPS file scaled to
%% 80% of its natural size with \epsscale. Its caption
%% has been written to indicate that additional figure parts will be
%% available in the electronic journal.

\begin{figure}
\epsscale{1.00}
\plotone{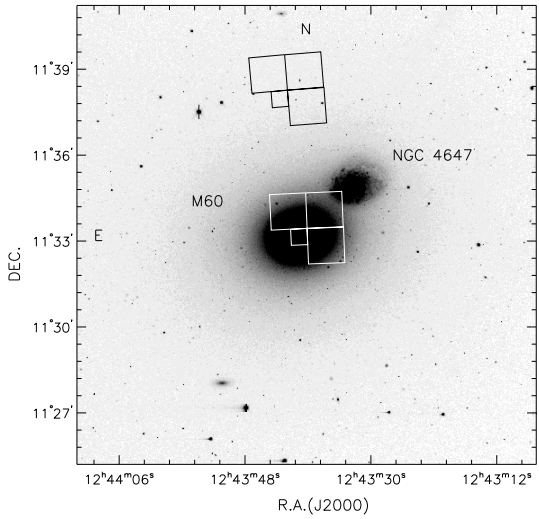} %wcsm60rsh.ps} %finder.eps} 
\caption{ %\scriptsize 
A grayscale map of the Washington $T_1$ image of M60.
The size of the field of view is $16\arcmin.4 \times 16\arcmin.4$. North is up and
east to the left.
%The halo light of M60 was subtracted from the original image to show better the point sources. %by IRAF/ELLIPSE fitting. 
The small spiral galaxy in the north-west ($\Delta$RA$=-109\arcsec.9$ and
$\Delta$Dec$ = 107\arcsec.3$) of M60 is NGC 4647.
The HST/WFPC2 fields % covering the center of M60 is 
are also marked by boxes.
\label{fig1}}
\end{figure}
%\clearpage

\begin{figure}
\epsscale{1.00}
\plotone{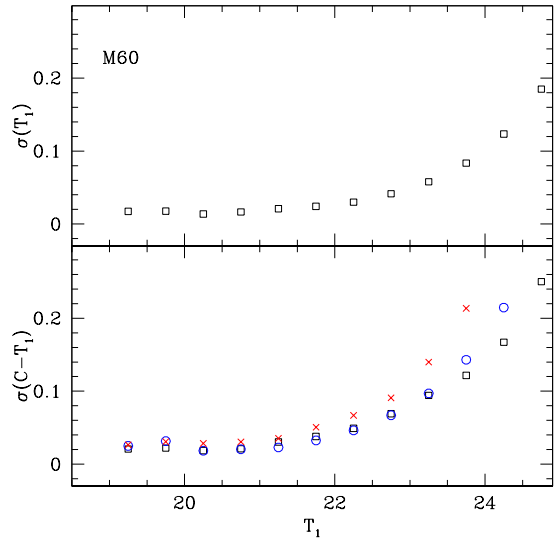} %Figerrnew.eps} 
\caption{ %\scriptsize 
Mean photometric errors of $T_1$ and ($C-T_1$) vs. $T_1$ magnitudes of the point sources measured in the long exposure images of M60.
In the low panel, squares, circles, and crosses represent the errors for all sources, 
blue globular clusters
($1.0<(C-T_1 ) <1.7$), and red globular clusters ($1.7<(C-T_1 ) <2.4$), respectively.
\label{fig2}}
\end{figure}
%\clearpage

\begin{figure}
\epsscale{1.00}
\plotone{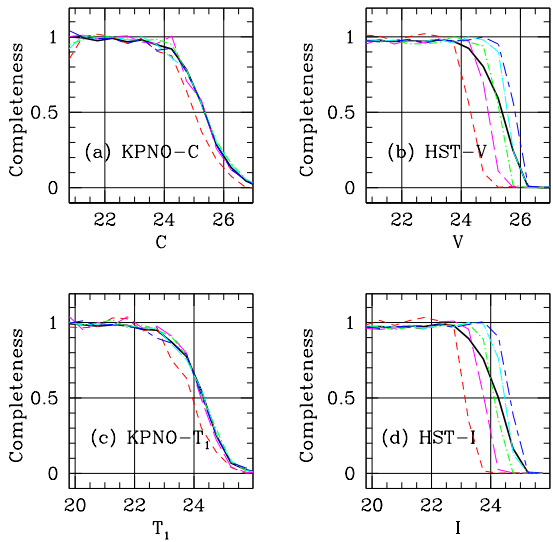} %Figerrnew.eps} 
\caption{ %\scriptsize 
Completeness for  KPNO $C$ (a) and $T_1$ (c): thick solid line for
the entire region, 
short dash line for $1\arcmin<r<2\arcmin$, long dash line for $2\arcmin<r<3\arcmin$, 
dot-short dash line for $3\arcmin<r<4\arcmin$, dot-long dash line for $4\arcmin<r<5\arcmin$, and
short dash-long dash line for $5\arcmin<r<6\arcmin$.
Completeness for  HST $V$ (b) and $I$ (d): thick solid line for
the entire region, 
short dash line for $r<10\arcsec$, long dash line for $10\arcsec<r<25\arcsec$, 
dot-short dash line for $25\arcsec<r<40\arcsec$, dot-long dash line for $40\arcsec<r<60\arcsec$, and
short dash-long dash line for $60\arcsec<r<80\arcsec$.
% and ???????? for $80\arcsec<r<100\arcsec$. 
\label{fig3}}
\end{figure}
%\clearpage
 
\begin{figure}
\epsscale{1.00}
\plotone{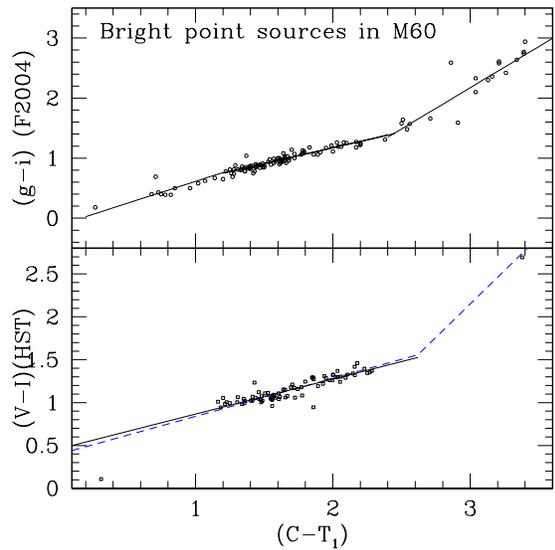} %Figcompgivi.eps} 
\caption{ %\scriptsize 
(a) \citet{for04}'s $(g-i)$ vs. ($C-T_1$) of the measured objects with $T_1<22$ mag in M60 in common between \citet{for04} and this study. 
The solid lines represent multi-linear fits to the data.
(b) HST $(V-I)$ vs. ($C-T_1$) of the measured objects with $V<22$ mag in M60. 
The solid line represents a linear fit to the data, and the dashed line represents
the fit derived from the photometry of M49 (NGC 4472) by Lee \& Kim (2000).
\label{fig4}}
\end{figure}
%\clearpage

\begin{figure}
\epsscale{1.00}
\plotone{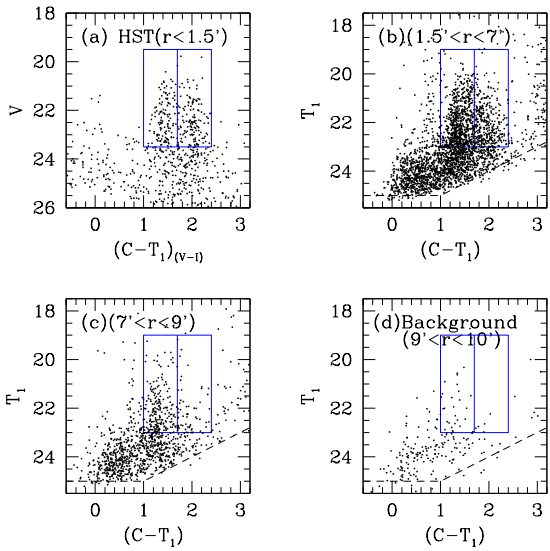} %Figcmdr.eps} 
\caption{ %\scriptsize  $T_1$--($C-T_1$) 
Color-magnitude diagrams of the point sources in M60. 
The dashed line represents an approximate lower boundary of the KPNO photometry.
The boxes represent the regions for selecting BGCs and RGCs with $19<T_1<23$ mag.
(a) $r<1\arcmin.5$ in the HST WFPC2 images. $(C-T_1)_{(V-I)}$ is the color
converted from $(V-I)$.
(b) $1\arcmin.5<r<7\arcmin$ except for the circular region with radius of one arcmin 
centered on NGC 4647. 
(c) $7\arcmin<r<9\arcmin$. % and 
(d) Background ($r>9\arcmin$). 
\label{fig5}}
\end{figure}
%\clearpage

\begin{figure}
\epsscale{1.00}
\plotone{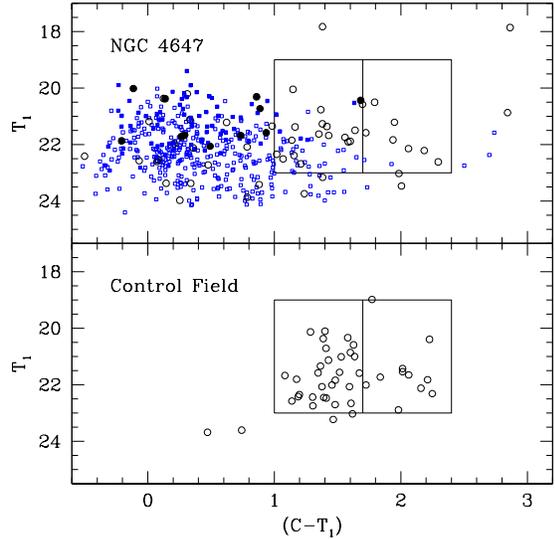} %Fign4647cmd.eps} 
\caption{ %\scriptsize 
$T_1$--($C-T_1$) color-magnitude diagrams of the point sources at 
$r<0\arcmin.5$ (filled circles) and $0\arcmin.5<r<1\arcmin$ (open circles) 
from the center of NGC 4647 (upper panel), and the point sources 
in the ``control field'' with the same area at the same
galactocentric distance from M60 as NGC 4647, but in the
opposite direction ($\Delta {\rm RA}=109.9\arcsec$ and
$\Delta {\rm Dec} = -107.3\arcsec$) (lower panel).
In the upper panel, filled squares and open squares represent slightly extended sources at
$r<0\arcmin.5$ and $0\arcmin.5 < r<1.0\arcmin$, respectively.
%These sources are mostly star clusters in NGC 4647.
Most of these sources are considered to be star clusters associated with NGC 4647.
The boxes are the same as in Fig. 4.
\label{fig6}}
\end{figure}
%\clearpage

\begin{figure}
\epsscale{1.00}
\plotone{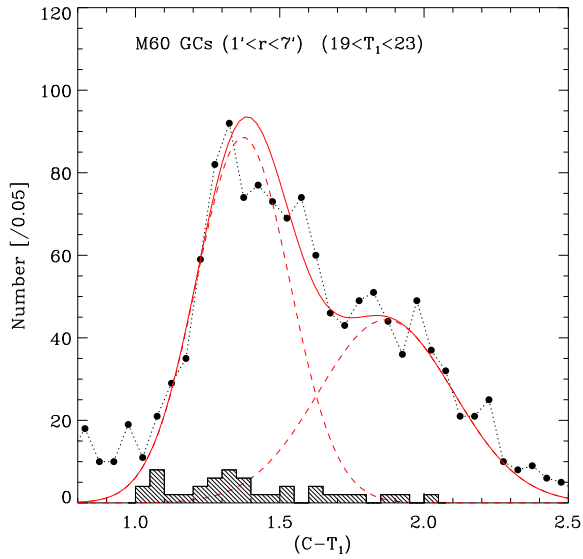} %m60gcct1hist2cmp.ps} %Figcdf.eps} %feh.eps} 
\caption{ %\scriptsize 
($C-T_1$) color distribution of the globular clusters with $19<T_1<23$ mag at $1\arcmin<r<7\arcmin$
from M60.
The thick solid line represents the double Gaussian fit to the data, each component
of which is plotted by the dashed lines.
The hatched histogram represents the color distribution of the background objects
at $9\arcmin<r<10\arcmin$.
%, which was subtracted from the original color distribution of the globular clusters.
\label{fig7}}
\end{figure}
%\clearpage

\begin{figure}
\epsscale{1.00}
\plotone{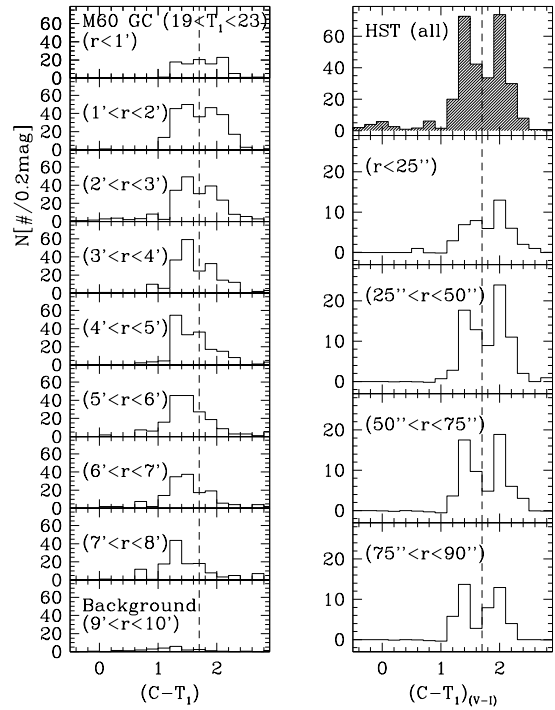} %Figcdfr.eps} 
\caption{ %\scriptsize 
Radial variation of the ($C-T_1$) color distribution of the globular clusters with $19<T_1<23$ mag in M60 after background subtraction.
(Left panel) KPNO photometry. (Right panel) HST/WFPC2 photometry.
\label{fig8}}
\end{figure}
%\clearpage

\begin{figure}
\epsscale{1.00}
\plotone{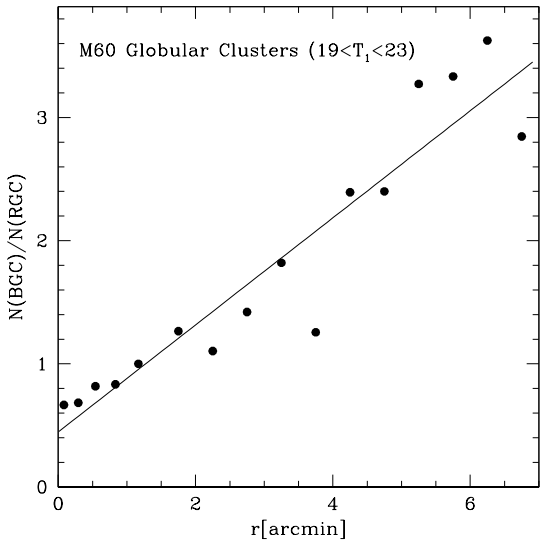} %Fignbr.eps} 
\caption{ %\scriptsize 
Radial variation of the number ratio of the BGCs and the RGCs
%blue globular clusters and the red globular clusters 
with $19<T_1<23$ mag in M60. The solid line represents a linear fit to the data.
\label{fig9}}
\end{figure}
%\clearpage

\begin{figure}
\epsscale{1.00}
\plotone{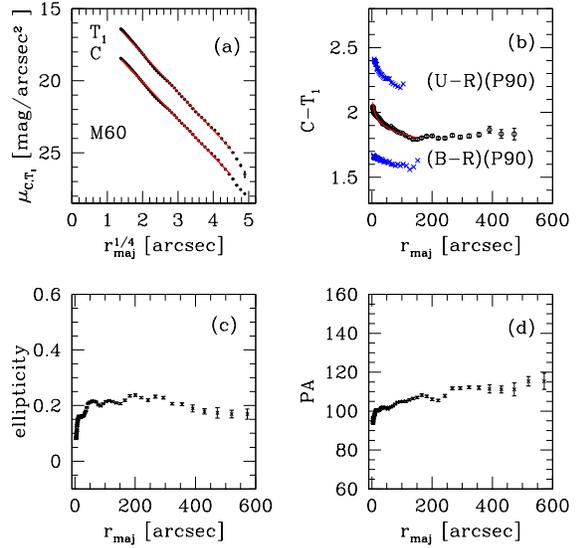} %m60surf1P90.eps m60surf1.eps} %Figsurf.eps} 
\caption{ %\scriptsize 
Surface photometry of M60. $r_{maj}$ represents galactocentric distance along the major axis.
(a) $C$ and $T_1$ surface brightness magnitudes vs. $r_{maj}^{1/4}$. The solid lines represent fits with $r^{1/4}$ law for $4\arcsec<r<410\arcsec$.
(b) Surface ($C-T_1$) color vs. $r_{maj}$.  
The solid line along the data represents a fit to the data: $(C-T_1 )=-0.162\pm 0.011 log r + 2.160$ 
for $3\arcsec<r<151\arcsec$ .
$(U-R)$ and $(B-R)$ colors given by \citet{pel90} are also plotted for comparison.
%$(C-T_1 )=-0.141\pm 0.018 log r + 2.137$  for $r<150\arcsec$ .
(c) Ellipticity  vs. $r_{maj}$.
(d) Position angle (PA) vs. $r_{maj}$. 
\label{fig10}}
\end{figure}
\clearpage

\begin{figure}
\epsscale{1.00}
\plotone{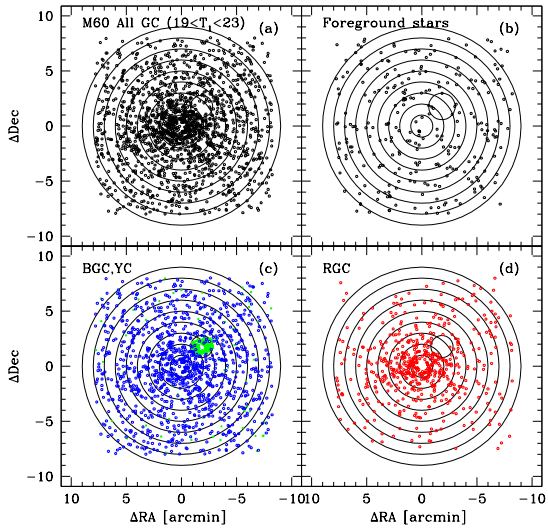} %Figrd4.eps} 
\caption{ %\scriptsize 
Spatial distribution of the globular clusters with $19<T_1<23$ mag.
The radii of the concentric circles are from 1 to 9 arcmin increasing by one arcmin,
and a small circle of radius 1 arcmin to the north-west of M60 is for NGC 4647.
(a) All globular clusters. % with $1.0<(C-T_1)<2.4$ .
(b) Foreground stars with $(C-T_1 )>2.5$ and $T_1<23$ mag. 
(c) BGCs %Blue globular clusters 
(open circles) and very blue objects
with $-0.5<(C-T_1 )<0.5$ (crosses, mostly in the region of NGC 4647).
(d) RGCs. %Red globular clusters. % with $1.7<(C-T_1)<2.4$.
\label{fig11}}
\end{figure}
%\clearpage

\begin{figure}
\epsscale{1.00}
\plotone{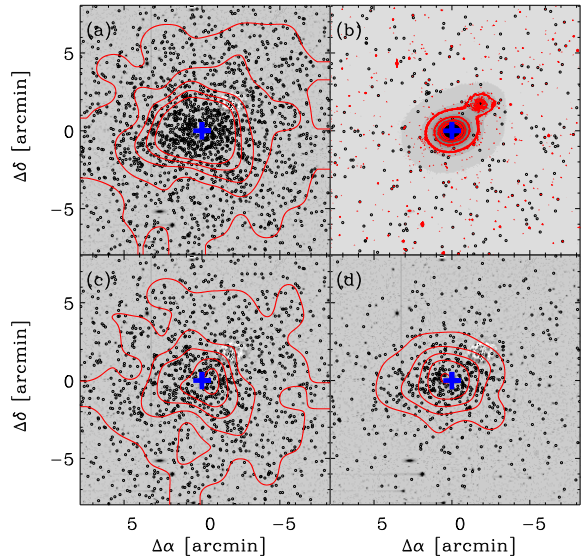} %m60spat.eps} %m60spatall.ps} %m60den4s.ps} %figdenkpno.eps} 
\caption{ %\scriptsize 
Number density contours of  the globular clusters with $19<T_1<23$ mag 
in comparison with the stellar halo of M60. The size of the field is $16\arcmin.4 \times16\arcmin.4$. 
Contour levels are the same in (a), (c) and (d):
%3.32344      6.64688      9.97033      13.2938      16.6172
3.323, 6.647, 9.970, 13.294, and 16.617 globular clusters per  arcmin$^2$.
The cross represents the center of M60.
(a) All globular clusters.
(b) A gray scale map of the $T_1$ image of M60, with isophotal contours.
Circles represent bright foreground stars.
% map overlayed on the greyscale map of M60 
(c) BGCs. %Blue globular clusters.
(d) RGCs. %ed globular clusters.
\label{fig12}}
\end{figure}
%\clearpage

\begin{figure}
\epsscale{1.00}
\plotone{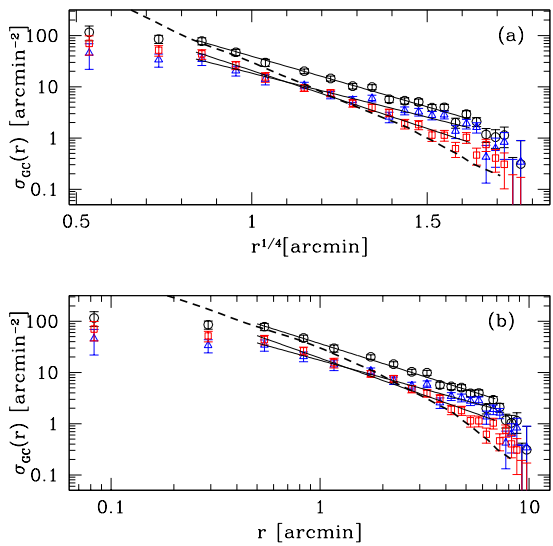} %figden2fit.eps} %Figden2.eps} %figdenkpno.eps} 
\caption{ %\scriptsize 
Number density profiles of  the globular clusters with $19<T_1<23$ mag 
(open circles for all globular clusters, open triangles for BGCs, %blue globular clusters, 
and open squares for RGCs). %red globular clusters).
Thick-dashed lines represent the $C$ surface brightness of
the stellar halo of M60, converted using $(26-\mu_C)/2.5$.
Solid lines represent fits to the three different datasets
 for $0\arcmin.5<r<7\arcmin$.
(a) Number density vs. $r^{1/4}$ [arcmin].
(b) Number density vs.  r [arcmin].
\label{fig13}}
\end{figure}
%\clearpage

\begin{figure}
\epsscale{1.00}
\plotone{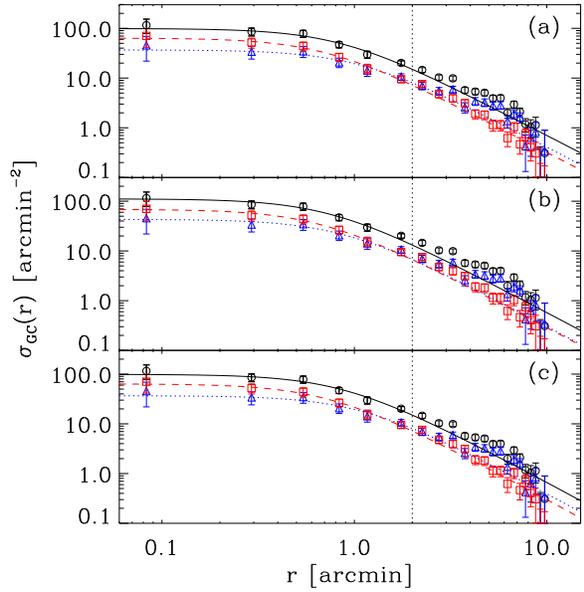} %numdenfit.ps %figden2fit.eps} %Figden2.eps} %figdenkpno.eps} 
\caption{ %\scriptsize 
Number density profiles of  the globular clusters with King model fits
%with $19<T_1<23$ mag 
(open circles for all globular clusters, open triangles for BGCs, %blue globular clusters, 
and open squares for RGCs). %red globular clusters).
%Thick-dashed lines represent the $C$ surface brightness of
%the stellar halo of M60, converted using $(26-\mu_C)/2.5$.
Curved lines represent King model fits to the three different datasets
 for $r<2\arcmin$ (marked by the vertical dotted line).
(a) Error weighted fit.
(b) Equal-weighted fit.
(c) Fit with the concentration parameter $c=2.5$.
\label{fig14}}
\end{figure}
%\clearpage

\begin{figure}
\epsscale{1.00}
\plotone{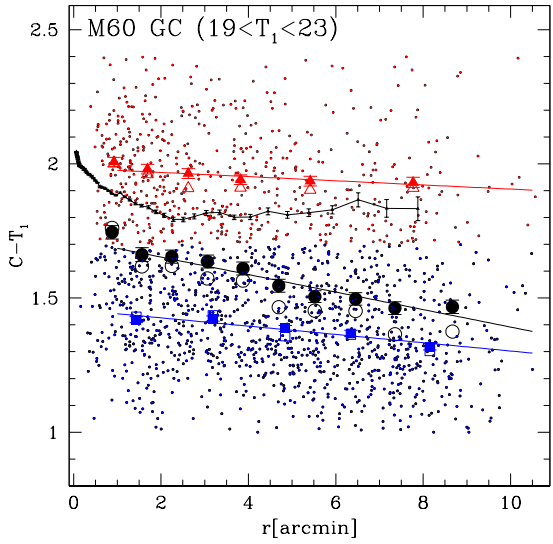} %figcolr1.eps} %Figcolr.eps} 
\caption{ %\scriptsize 
($C-T_1$) color vs. galactocentric distance for the globular clusters with $19<T_1<23$ mag (dots)
in comparison with the $(C-T_1 )$ surface color profile of stellar halo (curved line) in M60.
Larger filled and open symbols represent, respectively, the mean and median 
colors in a given radial                  bin, and
solid lines are linear fits to the mean colors.
Triangles, circles and squares represent, respectively, RGCs, %red globular clusters,
 all globular cluster, and BGCs. %blue globular clusters.
\label{fig15}}
\end{figure}
%\clearpage

\begin{figure}
\epsscale{1.00}
\plotone{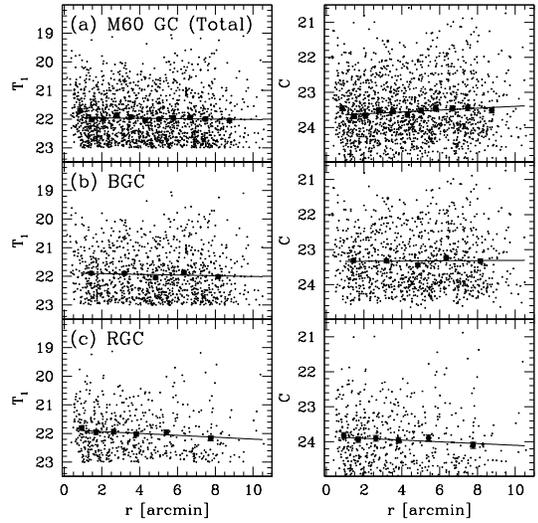} %Figmagr.eps} 
\caption{ %\scriptsize 
$T_1$ (left panel) and $C$ (right panel) magnitudes vs. galactocentric distance for the globular clusters with  $19<T_1<23$ mag in M60.
(a) All globular clusters.
(b) BGCs. %Blue globular clusters. % with $1.0<(C-T_1)<1.7$.
(c) RGCs. %Red globular clusters. % with $1.7<(C-T_1)<2.4$.
Filled circles represent the mean magnitudes in a given radial bin, and
solid lines are linear fits to the mean magnitudes.
\label{fig16}}
\end{figure}
%\clearpage

\begin{figure}
\epsscale{0.8}
\plotone{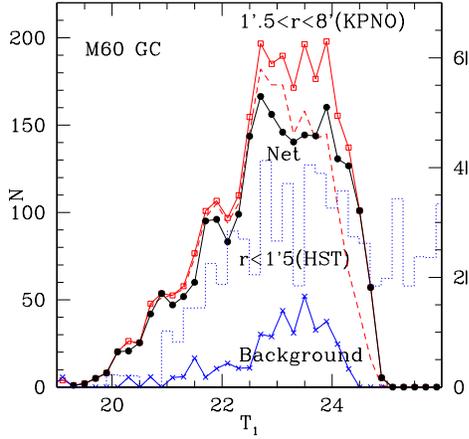} %figlfhstf.eps} %{Figlf.eps}
\caption{ Luminosity function of the globular clusters in M60:
dashed line for the raw luminosity function of the globular clusters at $1\arcmin.5<r<8\arcmin$ derived from the KPNO data,
open squares for the luminosity function corrected for incompleteness,
histogram for the luminosity function derived from the HST data
for $r<1\arcmin.5$, and 
cross for the luminosity function for the background objects with the same color
as the globular clusters at $9\arcmin<r<10\arcmin$ normalized to the area for $1\arcmin.5 <r< 8\arcmin$.
Filled circles represent the net luminosity function for $1\arcmin.5<r<8\arcmin$
after subtracting the contribution due to the background objects.
The HST luminosity function corrected for the completeness was scaled arbitrarily to match the KPNO luminosity function.
\label{fig17}}
\end{figure}
%\clearpage

\begin{figure}
\epsscale{1.00}
\plotone{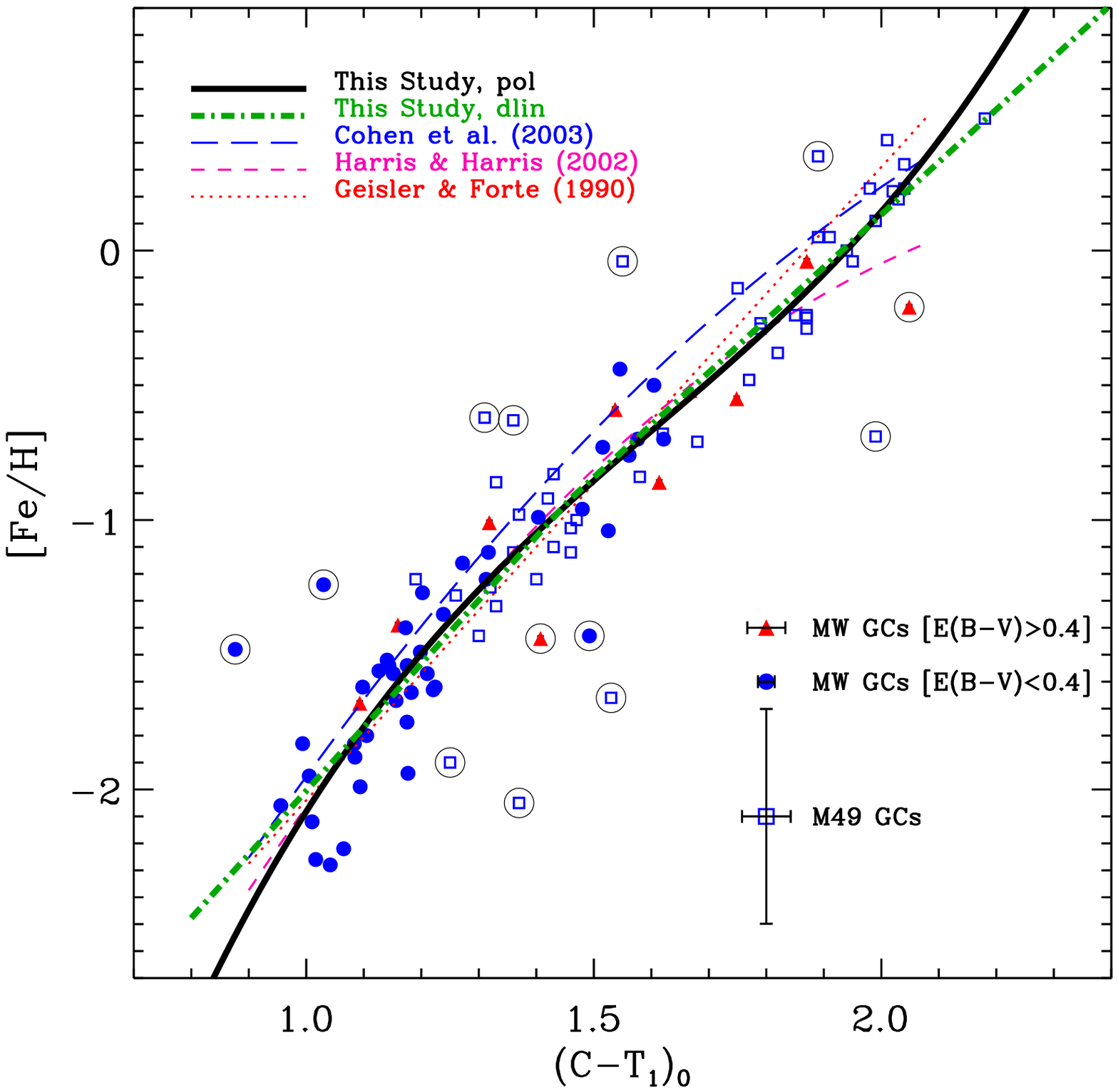} %ct1fehfitall.eps} %{ct1fehfitall.ps} %feh.eps} 
\caption{ %\scriptsize 
 [Fe/H] versus $(C-T_1)_0$ for the
globular clusters in our Galaxy (filled circles for $E(B-V)<0.4$, and
filled triangles for $E(B-V)>0.4$) and M49 (open squares).
Lines represent the transformation relations between  [Fe/H] and $(C-T_1)_0$ 
by \citet{gei90} %Geisler \& Forte (1990) 
(dotted line),
\citet{har02} %Harris \& Harris (2002) 
(short dashed curve),
\citet{coh03} %Cohen \etal (2003) 
(long dashed curve), and
double linear relations (thick dashed-dotted line) and 3rd order polynomial (thick solid curve) derived in this study.
Large open circles indicate the globular clusters that were not used for the derivation
of the transformation relations in this study.
Mean errorbars are plotted in the lower right region.
\label{fig18}}
\end{figure}
\clearpage

\begin{figure}
\epsscale{1.00}
\plotone{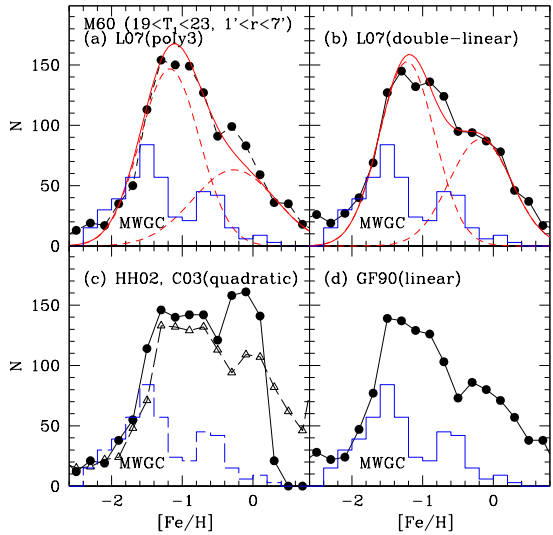} %Figcdffehnew4.eps} %figcdffehnew.eps} %Figcdffeh.eps} %feh.eps} 
\caption{ %\scriptsize 
 Metallicity [Fe/H] distribution of the globular clusters in M60 ($19<T_1 <23$ and $1\arcmin <r < 7\arcmin$) (circles), in comparison with the Galactic globular clusters (histogram).
[Fe/H] was derived from the $(C-T_1)_0$ color using the following
transformation relations:
3rd order polynomial equation (circles in (a)), double linear equation (circles in (b)) in this study,
and  \citet{har02} (circles in (c)), \citet{coh03} (triangles in (c)), and \citet{gei90} (circles in (d)).
In (a) and (b) smooth solid lines and dashed lines represent the Gaussian fits to the data. 
\label{fig19}}
\end{figure}
%\clearpage

\begin{figure}
\epsscale{1.00}
\plotone{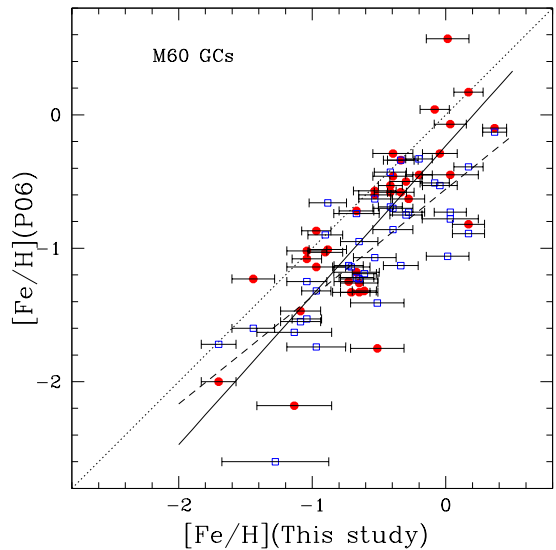} %m60fehfehp06.eps} %Figcdfm49.eps}
 \caption{Comparison of  spectroscopic [Fe/H] given by \citet{pie06} and photometric [Fe/H] derived using the double
linear relation in this study for 38 common globular clusters in M60.
Filled circles and open squares represent, respectively, [Fe/H] derived using the simple
stellar population (SSP) models and Brodie \& Huchra (BH) method by \citet{pie06}.
Solid line and dashed line represent linear fits to the data for SSP [Fe/H] and BH [Fe/H], respectively, and dotted line represents one-to-one correspondence.
\label{fig20}}
\end{figure}
%\clearpage

\begin{figure}
\epsscale{1.00}
\plotone{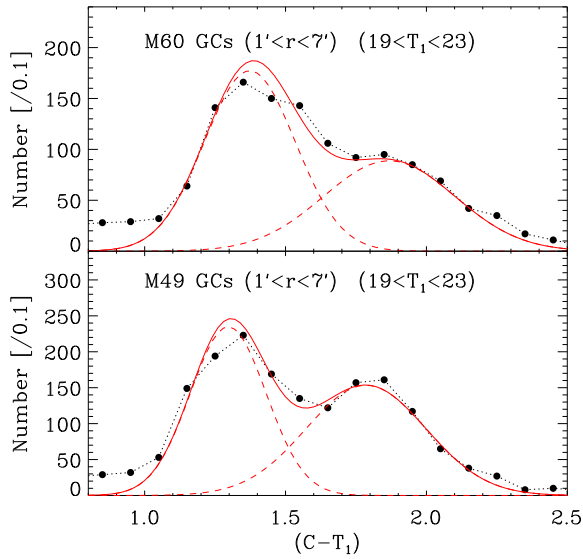} %m60m49col.eps} %Figcdfm49.eps}
 \caption{ %\scriptsize figncdfm49.eps
Comparison of the color distribution of the globular clusters
with $19<T_1<23$ mag and $1\arcmin<r<7\arcmin$
in M60 (this study, upper panel) and M 49 (Geisler, Lee, \& Kim 1996, lower panel).
Solid lines represent double Gaussian fits, while the dashed lines represent
the individual Gaussian components.
\label{fig21}}
\end{figure}
%\clearpage

\begin{figure}
\epsscale{1.0}
\plotone{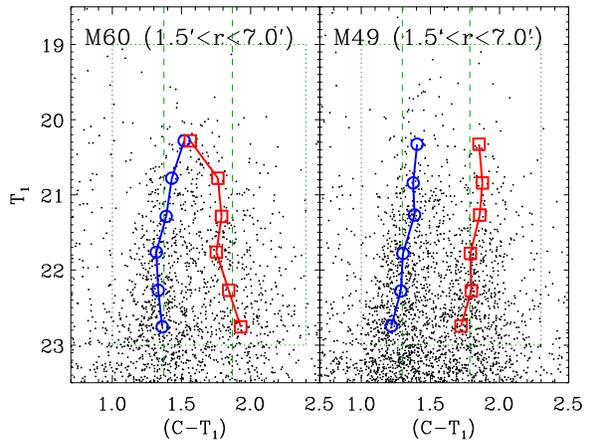} %btiltallfigtest3.eps} %btiltallfig.eps} %test1.ps} %figcmdrfitm60m49.eps} 
\caption{ %\scriptsize figncdfm49.eps
Comparison of the radial variation of the color magnitude diagrams of the globular clusters %with $r<7\arcmin$
in M60 (this study - left panel) and M 49 (Geisler, Lee, \& Kim 1996 - right panel).
The data are plotted for the objects at $1\arcmin.5 <r< 7\arcmin$ of both galaxies.  
%Top panels ($r<1\arcmin.5$) are from HST/WFPC2 photometry and lower panels ($1\arcmin.5<r<4\arcmin.5$, $4\arcmin.5 <r< 7\arcmin$, and $1\arcmin.5 <r< 7\arcmin$) 
%from KPNO photometry.
Dotted boxes represent the boundary for the globular clusters, and vertical dashed lines represent
the peak values for the BGCs and RGCs.
Circles and squares represent  the Gaussian centers for BGCs and RGCs, respectively,
derived for the magnitude bins using the KMM mixture model.
%Solid lines are linear fits to the data for the blue globular clusters 
%($20.5<T_1<22  $) marked by filled circles.
\label{fig22}}
\end{figure}

\end{document}